\documentclass[11pt,preprint]{aastex}

\def\eg{{\it e.g.}}
\def\ie{{\it i.e.}}
\def\kms{\ {\rm km\,s^{-1}}}
\def\hmpc{\, h^{-1}{\rm Mpc}}
\def\rs{r^*}
\def\us{u^*}
\def\gs{g^*}
\def\is{i^*}

\newcounter{thefigs}

\begin{document}
\title{Galaxy Clustering in Early SDSS Redshift Data} 

\author{Idit Zehavi\altaffilmark{1,2}, Michael R.\ Blanton\altaffilmark{1,3}, 
Joshua A.\ Frieman\altaffilmark{1,2}, David H.\ Weinberg\altaffilmark{4}, 
Houjun J.\ Mo\altaffilmark{5}, Michael A.\ Strauss\altaffilmark{6}, 
Scott F.\ Anderson\altaffilmark{7},   
James Annis\altaffilmark{1}, 
Neta A.\ Bahcall\altaffilmark{6},
Mariangela Bernardi\altaffilmark{2}, 
John W.\ Briggs\altaffilmark{8}, 
Jon Brinkmann\altaffilmark{9}, 
Scott Burles\altaffilmark{1}, 
Larry Carey\altaffilmark{7}, 
Francisco J.\ Castander\altaffilmark{2},
Andrew J.\ Connolly\altaffilmark{10}, 
Istvan Csabai\altaffilmark{11,12},
Julianne J.\ Dalcanton\altaffilmark{7},   
Scott Dodelson\altaffilmark{1,2},
Mamoru Doi\altaffilmark{13},
Daniel Eisenstein\altaffilmark{14},
Michael L.\ Evans\altaffilmark{7},  
Douglas P.\ Finkbeiner\altaffilmark{6}, 
Scott Friedman\altaffilmark{12}, 
Masataka Fukugita\altaffilmark{15}, 
James E.\ Gunn\altaffilmark{6},
Greg S.\ Hennessy\altaffilmark{16},
Robert B.\ Hindsley\altaffilmark{17}, 
\v{Z}eljko Ivezi\'{c}\altaffilmark{6},
Stephen Kent\altaffilmark{1,2},
Gillian R.\ Knapp\altaffilmark{6},
Richard Kron\altaffilmark{1,2}, 
Peter Kunszt\altaffilmark{18},
Donald Q.\ Lamb\altaffilmark{2,19},
R.\ French Leger\altaffilmark{1}, 
Daniel C.\ Long\altaffilmark{9}, 
Jon Loveday\altaffilmark{20}, 
Robert H.\ Lupton\altaffilmark{6},
Timothy McKay\altaffilmark{21},
Avery Meiksin\altaffilmark{22},
Aronne Merrelli\altaffilmark{23},
Jeffrey A.\ Munn\altaffilmark{15},  
Vijay Narayanan\altaffilmark{6},
Matt Newcomb\altaffilmark{24},
Robert C.\ Nichol\altaffilmark{24}, 
Russell Owen\altaffilmark{7},   
John Peoples\altaffilmark{1}, 
Adrian Pope\altaffilmark{24,12}, 
Constance M.\ Rockosi\altaffilmark{2}, 
David Schlegel\altaffilmark{6},
Donald P.\ Schneider\altaffilmark{25}, 
Roman Scoccimarro\altaffilmark{3},   
Ravi K.\ Sheth\altaffilmark{1}, 
Walter Siegmund\altaffilmark{7},  
Stephen Smee\altaffilmark{26}, 
Yehuda Snir\altaffilmark{24},  
Albert Stebbins\altaffilmark{1},
Christopher Stoughton\altaffilmark{1}, 
Mark SubbaRao\altaffilmark{2}, 
Alexander S.\ Szalay\altaffilmark{12},
Istvan Szapudi\altaffilmark{27},
Max Tegmark\altaffilmark{28},
Douglas L.\ Tucker\altaffilmark{1},
Alan Uomoto\altaffilmark{12}, 
Dan Vanden Berk\altaffilmark{1},
Michael S.\ Vogeley\altaffilmark{29},
Patrick Waddell\altaffilmark{7},     
Brian Yanny\altaffilmark{1}, 
and Donald G.\ York\altaffilmark{2,19}, 
for the SDSS Collaboration
}

\altaffiltext{1}{Fermi National Accelerator Laboratory, P.O.\ Box 500, Batavia,
IL 60510, USA}
\altaffiltext{2}{Astronomy and Astrophysics Department, University of
Chicago, Chicago, IL 60637, USA}
\altaffiltext{3}{Department of Physics, New York University, 4 Washington
Place, New York, NY 10003, USA}
\altaffiltext{4}{Department of Astronomy, Ohio State University, 
Columbus, OH 43210, USA}
\altaffiltext{5}{Max-Planck-Institute for Astrophysics, 
Karl-Schwarzschild-Strasse 1, D-85741 Garching, Germany}
\altaffiltext{6}{Princeton University Observatory, Peyton Hall, Princeton, 
NJ 08544, USA}
\altaffiltext{7}{Department of Astronomy, University of Washington, 
Box 351580, Seattle, WA 98195, USA}
\altaffiltext{8}{Yerkes Observatory, University of Chicago, 373 West
Geneva Street, Williams Bay, WI 53191, USA}
\altaffiltext{9}{Apache Point Observatory, P.O.\ Box 59, Sunspot, NM 88349, 
USA}
\altaffiltext{10}{University of Pittsburgh, Department of Physics and
Astronomy, 3941 O'Hara Street, Pittsburgh, PA 15260,USA}
\altaffiltext{11}{Department of Physcis, E\"otv\"os University, Budapest, 
Pf.\ 32, Hungary, H-1518}
\altaffiltext{12}{Department of Physics and Astronomy, The Johns Hopkins
University, 3701 San Martin Drive, Baltimore, MD 21218, USA}
\altaffiltext{13}{Institute of Astronomy and Research Center for the Early
Universe, School of Science, University of Tokyo, 113-0033, Japan}
\altaffiltext{14}{Steward Observatory, University of Arizona, 933 N.\ 
Cherry Ave., Tucson, AZ 85721, USA}
\altaffiltext{15}{Institute for Cosmic Ray Research, University of Tokyo, 
Kashiwa 277-8582, Japan}
\altaffiltext{16}{U.S.\ Naval Observatory, 3450 Massachusetts Ave., NW,
Washington, DC, 20392, USA}
\altaffiltext{17}{Remote Sensing Division, Code 7210, Naval Research 
Laboratory, 4555 Overlook Ave., SW, Washington, DC 20375, USA}
\altaffiltext{18}{CERN, IT Division, 1211 Geneva 23, Switzerland}
\altaffiltext{19}{Enrico Fermi Institute, University of Chicago, Chicago, 
IL 60637, USA}
\altaffiltext{20}{Sussex Astronomy Centre, University of Sussex, Falmer,
Brighton BN1 9QJ, UK}
\altaffiltext{21}{Department on Physics, University of Michigan, Ann Arbor,
MI 48109, USA}
\altaffiltext{22}{Department of Physics \& Astronomy, The 
University of Edinburgh, James Clerk Maxwell Building, The King's 
Buildings, Mayfield Road, Edinburgh, EH9 3JZ, UK}
\altaffiltext{23}{Department of Astronomy, California Institute of 
Technology, Pasadena, CA 91125, USA}
\altaffiltext{24}{Department of Physics, 5000 Forbes Avenue, Carnegie
Mellon University, Pittsburgh, PA 15213, USA}
\altaffiltext{25}{Department of Astronomy and Astrophysics, The Pennsylvania
State University, University Park, PA 16802,USA}
\altaffiltext{26}{Department of Astronomy, University of Maryland, College
Park, MD 20742, USA}
\altaffiltext{27}{Institute for Astronomy, University of Hawaii, 2680
Woodlawn Drive, Honolulu, HI 96822, USA}
\altaffiltext{28}{Department of Physics, University of Pennsylvania,
Philadelphia, PA 19101, USA}
\altaffiltext{29}{Department of Physics, Drexel University, Philadelphia,
PA 19104, USA}

\begin{abstract}

We present the first measurements of clustering in the Sloan Digital Sky Survey
(SDSS) galaxy redshift survey.  Our sample consists of $29,300$ galaxies
with redshifts $5,700\kms \leq cz \leq 39,000\kms$, distributed in several
long but narrow ($2.5-5^\circ$) segments, covering 690 square degrees. 
For the full, flux-limited sample, the redshift-space correlation length 
is approximately $8\hmpc$.  The two-dimensional 
correlation function $\xi(r_p,\pi)$ shows clear signatures of both the 
small-scale, ``fingers-of-God'' distortion caused by velocity
dispersions in collapsed objects and the large-scale compression caused by
coherent flows, though the latter cannot be measured with high precision
in the present sample.  The inferred real-space correlation function is well 
described by a power law, $\xi(r)=(r/6.1\pm0.2\hmpc)^{-1.75\pm0.03}$, for 
$0.1\hmpc\leq r\leq 16\hmpc$.  The galaxy pairwise velocity dispersion 
is $\sigma_{12}\approx 600\pm100\kms$ for projected separations 
$0.15\hmpc \leq r_p \leq 5\hmpc$.  When we divide the sample by color, the 
red galaxies exhibit a stronger and steeper real-space correlation function 
and a higher pairwise velocity dispersion than do 
the blue galaxies.  The relative behavior of subsamples defined by high/low 
profile concentration or high/low surface brightness is qualitatively similar 
to that of the red/blue subsamples.  Our most striking result is a clear 
measurement of 
scale-independent luminosity bias at $r \la 10\hmpc$: subsamples with absolute 
magnitude ranges centered on $M_*-1.5$, $M_*$, and $M_*+1.5$ have real-space 
correlation functions that are parallel power laws of slope $\approx -1.8$ 
with correlation lengths of approximately $7.4\hmpc$, $6.3\hmpc$, and 
$4.7\hmpc$, respectively. 
\end{abstract}




\section{Introduction}
\label{sec:intro}

The primary observational goals of the Sloan Digital Sky Survey (SDSS)
are to image 10,000 square degrees of the North Galactic Cap in five
passbands, with an $r$-band limiting magnitude of 22.5, to obtain
spectroscopic redshifts of $10^6$ galaxies and $10^5$ quasars,
and to obtain similar data for three $\sim 200$ square
degree stripes in the South Galactic Cap, with repeated imaging to enable
co-addition and variability studies in one of these stripes \citep{york00}.
One of the principal scientific objectives is to map
the large-scale structure traced by optical galaxies with unprecedented
precision over a wide range of scales.  These measurements of 
large-scale structure will allow critical tests of cosmological models
and theories of galaxy formation.  This paper presents the first 
measurements of galaxy clustering from the SDSS redshift survey, based
on a sample of $\sim 30,000$ galaxies observed during commissioning
operations and during the first few months of the survey proper.
Complementary studies of the angular clustering of galaxies in the
SDSS imaging survey appear in \citet{connolly01} and \citet{tegmark01},
and the implications of these measurements for the 3-D galaxy power spectrum
are discussed by \citet{dodelson01} and \citet{szalay01}.  \citet{scranton01}
examine many possible systematic effects on the angular clustering 
measurements and conclude that they are small; these tests and conclusions
are also relevant to the analyses of the redshift survey carried out here.

The redshift-space clustering of galaxies has been a central concern of
observational cosmology since the early studies of \citet{gregory78}
and \citet{joeveer78}.  Milestones in this effort include:
the first CfA redshift survey \citep{huchra83}, which mapped $\sim 2400$
galaxies selected from the \citet{zwicky68} catalog over 2.7 sr of sky 
to a magnitude limit of $m_{\rm Zw}=14.5$;
the Perseus-Pisces survey \citep{giovanelli85} consisting of $\sim 5000$ 
galaxies chosen from the CGCG and UGC catalogs; 
redshift surveys in other areas of sky such as the 
Southern Sky Redshift Survey \citep{dacosta91}
and the Optical Redshift Survey \citep{santiago95};
sparsely sampled surveys of optically selected galaxies to
$B \approx 17$ (the Stromlo-APM Redshift Survey, \citealt{loveday96};
the Durham/UKST Redshift Survey, \citealt{ratcliffe98}); 
the second CfA redshift survey \citep{delapparent86,geller89},
with a magnitude limit of $m_{\rm Zw}=15.5$ and an eventual total of 
$\sim 13,000$
galaxies in the ``Updated Zwicky Catalog'' \citep{falco99};
a similar extension of the Southern Sky Redshift Survey \citep{dacosta98};
redshift surveys of IRAS-selected galaxies to successively deeper
flux limits of 2 Jy \citep{strauss92}, 1.2 Jy \citep{fisher95},
and 0.6 Jy (the sparsely sampled QDOT survey, \citealt{lawrence99}; 
the PSCz survey of $\sim 15,000$ galaxies, \citealt{saunders01}); 
the deep slice surveys of Vettolani et al. (\citeyear{vettolani98};
$\sim 3300$ galaxies to $b_J=19.4$) and Geller et al. (\citeyear{geller97};
$\sim 1800$ galaxies to $R=16.13$);
and the Las Campanas Redshift Survey (LCRS; \citealt{shectman96}),
which mapped $\sim 24,000$ galaxies in six thin ($1.5^\circ \times 90^\circ$)
slices at a depth $R \approx 18$.  The current state-of-the-art is represented
by \citet{peacock01} and Percival et al.'s (\citeyear{percival01}) 
studies of redshift-space clustering in a sample of $\sim 140,000$ galaxies 
from the ongoing 2dF Galaxy Redshift Survey (2dFGRS).
The sample that we analyze here is most similar to the LCRS,
with slightly more galaxies but a comparable depth and thin-slice geometry.

Two factors that complicate and enrich the interpretation of galaxy
clustering in redshift surveys are the distortions of structure induced
by peculiar velocities and the possibility that galaxies are
``biased'' tracers of the underlying matter distribution.
On small scales, velocity dispersions in collapsed objects 
(a.k.a. ``fingers-of-God'') smear out structures along the line of sight,
effectively convolving the real-space correlation function with the galaxy 
pairwise velocity distribution (see, \eg, \citealt{davis83}).  On large scales,
coherent flows into high density regions and out from low density
regions enhance structures along the line of sight 
\citep{sargent77,kaiser87,regos91,weygaert93,hamilton98}. 
Because the underlying clustering pattern should be statistically
isotropic, the apparent anisotropy induced by redshift-space distortions
yields constraints on the distribution of peculiar velocities, which
can in turn yield constraints on the matter density parameter $\Omega_m$.
With our current galaxy sample, we clearly detect the signature of both
the small-scale, ``fingers-of-God'' suppression and the large-scale,
coherent flow amplification.  However, we are not yet able to measure
the latter effect with high precision, so we defer a detailed examination
of $\Omega_m$ constraints (and comparison to \citealt{peacock01}) 
to a future analysis of a larger sample.

The notion that the optical galaxy population might give a systematically 
``biased'' picture of matter clustering came to the fore in the mid-1980s, 
largely in an effort to reconcile the predictions of $\Omega_m=1$ 
inflationary models with observations 
\citep{davis85,bardeen86,melott86,bahcall83,kaiser84}.  
There are now numerous arguments in favor of a low-$\Omega_m$
universe, but theoretical models of galaxy formation, the well known
dependence of observed galaxy clustering on morphological type 
(\eg, \citealt{hubble36,zwicky37,abell58,davis76,dressler80,guzzo97}),
and more recent evidence for dependence of clustering on luminosity
(\eg, \citealt{hamilton88,white88,park94,loveday95,benoist98,willmer98}) 
all imply that galaxies cannot be perfect tracers of the underlying
matter distribution.  Advances in hydrodynamic cosmological
simulations, high-resolution N-body simulations, and semi-analytic methods
now allow detailed {\it a priori} predictions of bias for physically
motivated models of galaxy formation (\eg, \citealt{cen92,katz92,
benson99,blanton99,colin99,kauffmann99,pearce99,white01,yoshikawa01}).
Empirical constraints on bias can therefore provide tests of galaxy
formation theories and guidance to physical ingredients that may be
missing from current models.  The SDSS is ideally suited to the empirical
study of bias because of the high sampling density and the detailed 
photometric and spectroscopic
information available for every galaxy.  We begin the effort here,
by examining the dependence of the real-space correlation function and the
redshift-space distortions on galaxy color, luminosity, surface brightness,
and light profile concentration.

The next Section describes the data sample used for the clustering
analysis.  Section 3 describes our methods for estimating the correlation
function, including technical issues such as sampling corrections and
the effects of the minimum fiber spacing in the spectroscopic observations.
Section 4 presents the clustering results for the full, flux-limited
galaxy sample.  Section 5 examines the clustering of subsamples defined
by color, luminosity, and other galaxy properties. We summarize our results 
in Section 6.  A discussion of our jackknife error estimation procedure, and 
comparison of this procedure to results from mock redshift catalogs, appears 
in Appendix A.  Throughout the paper, absolute magnitudes quoted for galaxies
assume $H_0=100\kms\;{\rm Mpc}^{-1}$.

\section{Data}
\label{sec:data}

\subsection{Description of the Survey}
\label{subsec:survey}

The SDSS (\citealt{york00}) is producing an imaging and spectroscopic
survey over $\pi$ steradians in the Northern Galactic Cap. A
dedicated 2.5m telescope (\citealt{siegmund01}) at
Apache Point Observatory, Sunspot, New Mexico, images the sky in five
bands between 3,000 and 10,000 \AA ($u$, $g$, $r$, $i$, $z$;
\citealt{fukugita96}) using a drift-scanning, mosaic CCD camera
(\citealt{gunn98}), detecting objects to a flux limit of $r\approx
22.5$.  Approximately 900,000 galaxies (down to
$r_{\mathrm{lim}}\approx 17.77$; \citealt{strauss01}), 100,000
Luminous Red Galaxies (LRGs; \citealt{eisenstein01}), and 100,000 quasars 
(\citealt{richards01}) are
targeted for spectroscopic follow up using two double fiber-fed 
spectrographs on the same telescope. Most of the essential technical details 
are summarized in a paper that accompanies the SDSS Early Data Release
(\citealt{stoughton01}).

As of June 2001, the SDSS has imaged around 2,500 square degrees of
sky and taken spectra of approximately 140,000 objects. We use a
subset of these data here to calculate the correlation function of
galaxies, confining our attention to regions where the data reductions
and calibration have been carefully checked and the spectroscopic
completeness is well understood.

\subsection{Imaging and Spectroscopic Pipelines}
\label{subsec:pipelines}

As described by \citet{stoughton01}, the imaging data are processed by
astrometric (\citealt{pier01}) and photometric (\citealt{lupton01a},
\citeyear{lupton01}) pipelines and calibrated relative to a
set of standard stars (\citealt{tucker01}).  Targets are selected by a
target selection pipeline (\citealt{vandenberk01}), and plates for
spectroscopic observations are drilled based on the results of a
tiling pipeline (\citealt{blanton01b}). After the spectra are
observed, the spectroscopic pipeline then reduces,
calibrates, and classifies the spectra and determines redshifts.

The photometric pipeline (\citealt{lupton01}) detects objects and
measures their properties in all five bands. Most relevant here are
the Petrosian magnitude $m_P$, the radius $r_{50}$
containing 50\% of the Petrosian flux, and the radius $r_{90}$
containing 90\% of the Petrosian flux.  The details of SDSS Petrosian
magnitudes, a modified form of those introduced by \citet{petrosian76}, 
are described in a number of references and will not be
repeated here, except to say that they are designed to 
measure a constant (and large) fraction of a galaxy's total light,
independent of redshift or central surface brightness but (slightly)
dependent on light-profile shape (\citealt{blanton01a};
\citealt{lupton01}; \citealt{strauss01}; \citealt{stoughton01}; 
\citealt{yasuda01}).  The radii $r_{50}$ and $r_{90}$, which we use below to 
quantify galaxies'
surface brightnesses and morphologies, are not corrected for
seeing. However, such corrections would be small
since most of the galaxies in this sample
are relatively large ($r_{50} > 2''$; \citealt{blanton01a}), and the
seeing conditions for the imaging are generally good (FWHM$\la 1.5''$).

The flux calibration is performed relative to standard stars as described
in \citet{york00}, \citet{tucker01} and \citet{stoughton01}. Calibration 
is a three-tiered system in which ``secondary standards'' that are not
saturated in the 2.5m imaging camera are used to calibrate the imaging
data. These secondary standards are themselves calibrated relative to
a set of ``primary standards'' using a 0.5m photometric telescope (PT;
\citealt{uomoto01}). These primary standards have been calibrated
relative to the fundamental standard BD $+ 17^\circ 4708$ by the
United States Naval Observatory 1m telescope. The calibrations used
here are not fully validated, though they are thought to be accurate
to within 5\%. Because of this remaining uncertainty, object magnitudes
are referred to in this paper and others based on early SDSS data as
$u^\ast$, $g^\ast$, $r^\ast$, $i^\ast$, and $z^\ast$.

The target selection pipeline (\citealt{vandenberk01}) determines
which objects from the imaging survey are spectroscopic targets. 
We concentrate here on the ``Main Sample'' galaxies in the SDSS, 
which are selected using the criteria detailed by \citet{strauss01}.
The essential selection criteria for this sample are the star--galaxy
separator, the surface-brightness limit, and the flux limit. The
star--galaxy separation is based on a comparison of the flux of the 
object measured through a point-spread function aperture to the flux 
estimated using a best-fit model to the galaxy profile (choosing the 
better of pure exponential and de Vaucouleurs profiles).  This method 
is known to be an extremely efficient and reliable separator at the 
magnitudes appropriate for the spectroscopic sample (\citealt{lupton01}). 
We find that $98\%$ of objects targeted as main sample galaxies indeed 
turn out to be galaxies.  The  major contaminant is double stars with 
separations less than $2''$. 

The surface-brightness limit is based on the Petrosian half-light
surface brightness in $r^\ast$.  For some parts of the sample used here,
obtained during commissioning observations, the surface-brightness limit 
is $\mu_{1/2}=23.5$ mag arcsec$^{-2}$, but for most of the sample 
it is $\mu_{1/2}=24.5$ mag arcsec$^{-2}$.
Because we will use relatively luminous galaxies to trace the
density field here, the positive correlation between
surface brightness and luminosity (\citealt{blanton01a}) guarantees
that the surface-brightness limit will be unimportant.

The flux limit of the spectroscopic survey is approximately
$r^{\ast}=17.77$, after correction for Galactic reddening using
the maps of Schlegel, Finkbeiner \& Davis (1998).
The limit varies somewhat over the area of our sample, as the target
selection criteria changed during the commissioning phase of the survey,
when much of these data were taken. We will cut back to a uniform flux 
limit of $r^\ast=17.6$ for our current
analysis. In addition, there is a bright limit imposed on the flux in
a 3-arcsecond diameter aperture (the entrance aperture of a spectroscopic
fiber) of $m_{\mathrm{fiber}}>15$ in $\gs$, $\rs$ and 
$m_{\mathrm{fiber}}>14.5$ in $\is$, in order
to avoid saturation and cross-talk between fibers in the spectrograph.

The reliability of the galaxy target selection is very high; galaxy
target selection results for two imaging runs over the same patch of
sky agree for 95\% of the objects; the differences are
attributable to small, random magnitude errors shifting objects across
the flux limit \citep{strauss01}.

The tiling pipeline (\citealt{blanton01b}) positions spectroscopic
tiles and assigns fibers to targets. The most important constraint is
imposed by the size of the fiber plugs, which dictates that two fibers
cannot be placed closer than $55''$ to each other. If the
spectroscopic tiles did not overlap, this would mean that about 10\%
of the objects would be unobservable. Because the tiles are circular, 
about $30\%$ of the sky is actually covered by more than a single tile; 
in these regions, many of the objects lost due to collisions of fibers
can be recovered.
Note, however, that the tiles are positioned such that there are more
tiles in dense areas of sky; thus, the regions covered by tile
overlaps tend to be 5--10\% overdense compared to average. We will
describe in Section~\ref{sec:method} how we handle objects whose redshifts
are missing due to fiber collisions. 

Finally, the spectroscopic pipeline extracts, analyzes, 
and classifies the spectra, determining the spectral type, redshift, and 
other spectral information for each target. The success rate for classifying 
spectra and determining redshifts correctly
is very high ($> 99\%$) for main sample galaxy targets, based 
on a subsample of $\sim 20,000$ spectra examined by eye.  
The spectroscopic pipeline assigns an empirically calibrated 
confidence level to the redshift determination for each object; 
cutting out main sample galaxy redshifts with low confidence 
(CL$ < 75\%$) removes only 0.7\% of the objects from the sample, 
with a negligible effect on the clustering results below.

\subsection{Determining Positions, Luminosities, and Rest-frame Colors}
\label{subsec:cosmo}

The redshift of a galaxy is not a linear measure of an object's
distance at the moderate redshifts probed here (median $z\sim
0.1$), and the comoving distance of an object depends somewhat on the
cosmology assumed. Throughout this paper, we assume a
Friedmann-Lema\^{\i}tre metric with $\Omega_m=0.3$ and
$\Omega_\Lambda=0.7$. When we plot correlation functions versus
separation, we are always referring to the comoving separation,
transformed from km s$^{-1}$ separations using the standard formulas
as tabulated in, for example, \citet{hogg99}.

We also must account for cosmological effects when calculating the
absolute magnitudes from the apparent magnitude and the redshift using
the formula
\begin{equation}
M = m - \mathrm{DM}(z) - K(z) + 5 \log_{10} h,
\end{equation}
where DM$(z)$ is the bolometric distance modulus for the cosmology in
question (again, see \citealt{hogg99}), $K(z)$ is the $K$-correction,
and the Hubble constant is $H_0 = 100 h \kms$ Mpc$^{-1}$. Throughout
this paper, we use $h=1$ to compute absolute magnitudes, and we
quote distances in $\hmpc$.

The $K$-correction is necessary to account for the fact that the
system response in the observed frame corresponds to a narrower, bluer
rest-frame passband, depending on the redshift of the observed
object. In order to make an estimate of the $K$-correction, it is
therefore necessary to have an estimate of the spectral energy
distribution (SED) of each object. We can make a good estimate based
on the five-band photometry provided by SDSS. For each object, we find
the linear combination of the four SED templates of
\citet{coleman80}, as extended in the red and blue by
\citet{bolzonella00}, which best fits the photometry. We use the
resulting SED to estimate the $K$-corrections, assuming no evolution
of the SED. This method is similar to simply interpolating between 
passbands to infer a rest-frame flux, while also taking
advantage of what astronomers know
already about galaxy SEDs. These $K$-corrections are also useful to
determine the rest-frame colors of objects from their observed colors.
The details of our procedure, which are based on the photometric
redshift methods of \citet{csabai00}, will be described in a
forthcoming paper.

\subsection{Description of the Sample}
\label{subsec:sample}

\begin{figure}[p]
{\includegraphics{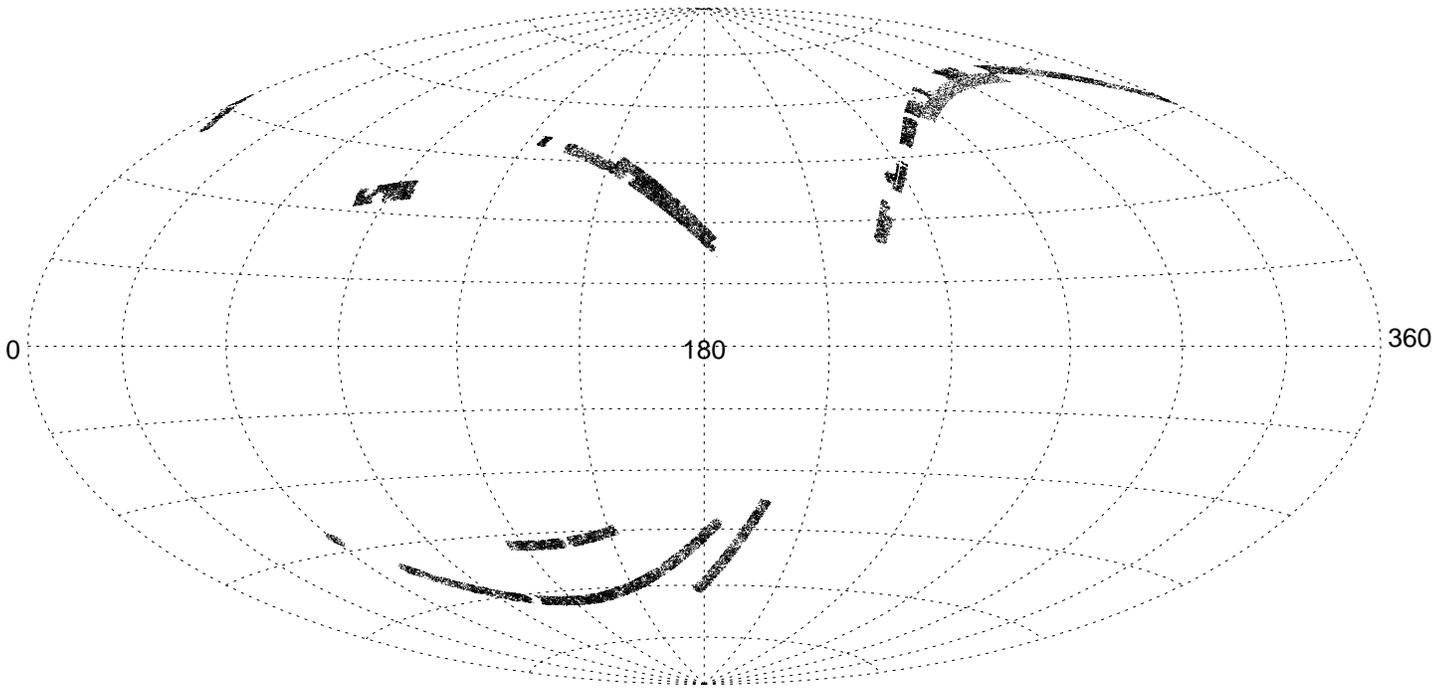}}
\vspace{13 cm}
\caption{Aitoff projection of our galaxy sample in Galactic coordinates.}
\label{fig:aitoff}
\end{figure}

\begin{figure}[tbp]
{\includegraphics{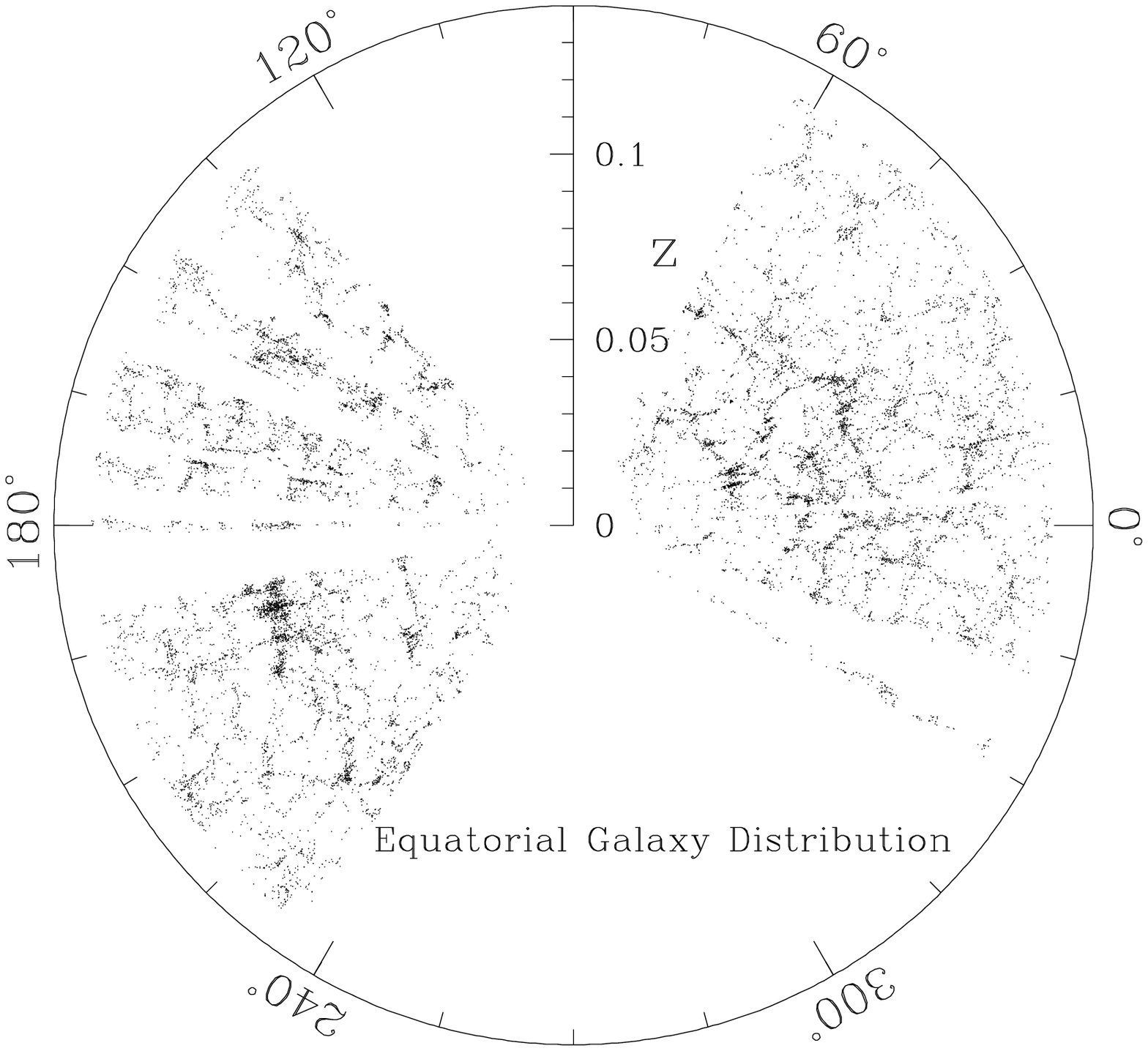}}
\vspace{16 cm}
\caption{Pie-diagram distribution for the equatorial part of our sample.
The plot includes $16,300$ galaxies that lie within $|\delta| <5^{\circ}$
of the Celestial Equator.
}
\label{fig:pie}
\end{figure}

Figure~\ref{fig:aitoff} shows the angular distribution of the
resulting sample in Galactic coordinates. The area covered is
approximately 690 square degrees (comparable to the sky coverage of the
LCRS survey), or about $7\%$ of the area that will
eventually be covered by the survey; in this area, we have selected
$\approx 30,000$ galaxies for our sample, as explained in the
following paragraphs. Figure~\ref{fig:pie} shows the distribution in
right ascension and redshift of galaxies near the Celestial Equator
($|\delta| < 5^\circ$).

Even though some regions of the survey are currently complete to
$r^{\ast} < 17.77$ (dereddened, using \citealt{schlegel98}), 
others are complete only to $r^\ast<17.6$, and for simplicity
we have pared back our sample to this constant flux limit.
In addition, we have imposed a bright limit of
$r^{\ast}>14.5$ because at the bright end we are limited by the bright
spectroscopic limits (the 3-arcsecond aperture magnitude limit of
$r^{\ast}>15$ imposed to prevent saturation and cross-talk of
fibers in the spectrograph) and by the quality of deblending of large
galaxies in the version of the photometric pipelines used for
targeting many of these galaxies. 
These flux limits reduce the number of targets we consider by about 10\%.

In most of this work, we limit our sample to a fairly small range in 
redshift, $5,700 \kms < cz < 39,000 \kms$.
We do so primarily because it is clear that galaxy
evolution within the full range of redshifts (which extends to about
$80,000 \kms$) is important, and at the time of this work there was not 
yet an adequate model of this evolution to allow proper calculation of the 
radial selection function. Working at low redshift primarily limits our 
estimate of the large-scale clustering; however, the thrust of this work is 
the small-scale clustering of galaxies. Much larger-area samples of SDSS
galaxies will soon be available, as well as good models of the
evolution of the luminosity function, and much better estimates of the
large-scale clustering will come from these samples. The outer redshift cut
is the most costly of our imposed limits, 
eliminating 30\% of the objects available after
the above flux limits have been imposed.

We wish to study the clustering of relatively luminous galaxies near the
exponential cutoff in the luminosity function at $M_\ast$.  For most
of the work below, we therefore impose absolute magnitude limits of
$-22 < M_{r^{\ast}} - 5 \log_{10} h< -19$, which roughly brackets the value
$M_\ast=-20.8$ determined for the SDSS (\citealt{blanton01a}). These
absolute magnitude limits exclude another 15\% of the objects (after
the redshift and flux cuts are imposed), leaving us with our canonical
sample of 29,300 galaxies. We will use slightly different cuts to
define volume limited samples of different luminosity ranges
below. Finally, we will compare below the clustering of several
different types of galaxies, defined by color, surface brightness, and
morphology, describing in the appropriate sections how those subsamples
of the canonical sample are defined.

\section{Measuring the Correlation Function}
\label{sec:method}

Before measuring the correlation function, we need to determine how
to treat the fiber collisions and how to properly correct for angular
and radial selection effects. We first detail how we account for these 
issues, then describe our estimators for the correlation function
and its errors.

\subsection{Accounting for Fiber Collisions}
\label{subsec:fiberc}

One of the important observational constraints in the SDSS is that no
two fibers on the same plate can be closer than $55''$. 
Thus, redshifts for both members of a close galaxy pair can only be obtained
in regions where tiles overlap.

If we took no account of fiber collisions at all, then we would
systematically underestimate correlations even on large scales because
collisions occur more often in overdense regions such as clusters,
which have enhanced large-scale clustering for the reasons discussed
by, {\it e.g.}, \citet{kaiser84}. A simple way to correct
this bias is to double-weight the member of each pair that {\it was}
observed, since its {\it a priori} selection probability was 50\%. Here 
we adopt a variant of the double-weighting procedure, assigning each 
pair member whose redshift was not obtained because of a fiber
collision the same redshift as the pair member whose redshift was measured.
We term such an assigned redshift a ``collision corrected'' redshift.  
On large scales, where both members of the pair contribute to the same
separation bin, the effect is the same as double weighting, but our
procedure should perform somewhat better on small scales because it
retains information about the known angular positions. Some of the galaxy 
targets are not assigned fibers due to collisions with QSOs or LRGs; in 
these cases, no redshift is assigned, and the galaxy is treated as if 
the fiber simply did not measure a redshift successfully, as described 
in the next subsection.

At $cz=39,000\kms$, the outer edge of our sample, $55''$ corresponds
to a comoving transverse separation of $0.1\hmpc$.  Fiber collisions
will have a significant effect on correlation function estimates below 
this scale, and in this paper we will restrict our measurements to 
separations $>0.1\hmpc$, so as to avoid the artificial increase of 
pairs with very small separations. Because two galaxies whose fibers
collide also have a line-of-sight separation,
the collisions can in principle affect our estimate
of the correlation function out to somewhat larger scales,
but we show below that these effects are probably smaller than our
statistical uncertainties for this sample.
In tile overlap regions, which constitute $\sim 30\%$ of the survey
area, we are able to measure redshifts for most of the galaxies which 
would otherwise be unobtainable due to fiber collisions.
We find that roughly $60\%$ of these galaxies in fact have a redshift
within $500\kms$ of their closest angular neighbor.
For the rest, either galaxy still has an equal 
chance of being selected, so double weighting does not statistically 
bias the correlations with more distant galaxies.  

\begin{figure}[tbp]
{\includegraphics{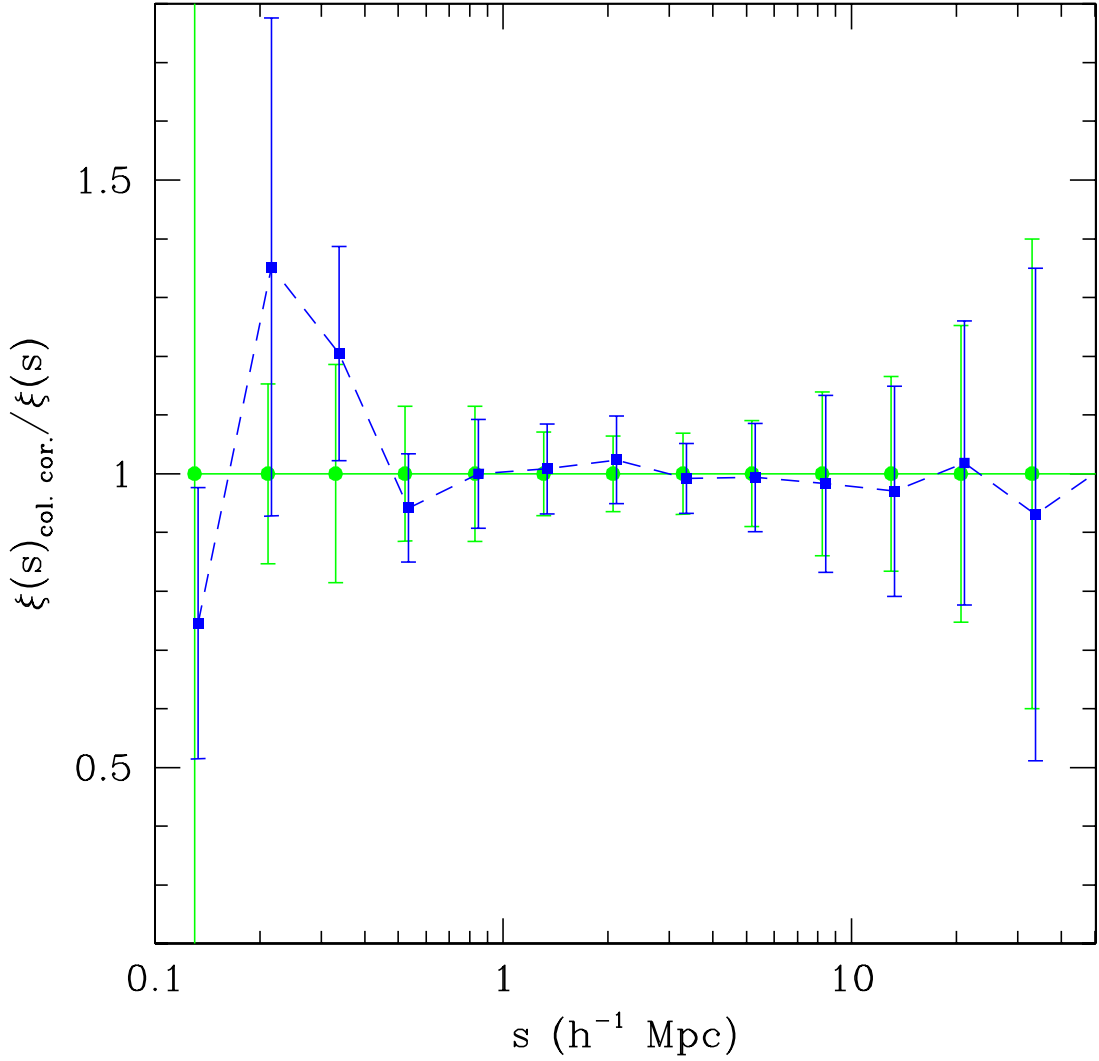}}
\vspace{16 cm}
\caption{Test of the accuracy of our correction for fiber collisions using
the tile overlap regions. The dashed line shows the ratio of the 
redshift-space correlation function measured using the ``collision corrected'' 
redshifts for all galaxies whose fibers collide to the result obtained 
using all the available observed redshifts. The latter estimate
(and errorbars) is given by the solid line at unity.
The points and errorbars are shifted slightly for clarity. }
\label{fig:fiberc}
\end{figure}

We can use the galaxies in the overlap regions (roughly a third of our 
full sample) to assess the accuracy of our correction procedure.  
For the standard analyses in this paper, we use the true redshifts of 
galaxies whose fibers collide whenever we can obtain them from overlapping 
tiles and the ``collision corrected'' redshifts from the closest angular 
neighbor when we cannot.  Figure~\ref{fig:fiberc} shows the results from 
the tile overlap regions when we instead use the ``collision corrected'' 
redshifts for all galaxies whose fibers collide.
We plot the ratio of the redshift-space correlation function obtained in
this way to the one obtained when we use all the available measured redshifts
(estimator and errors are defined in \ref{subsec:estimate}). 
The ``collision corrected'' redshifts yield a correlation function 
that is nearly identical to the one obtained from the observed redshifts
over a large range of scales. The two results deviate somewhat for small 
separations,  $s < 1\hmpc$, but they agree within the statistical 
uncertainties.
Note that these deviations are also partly statistical fluctuations due to
the smaller sample used for this comparison. In addition, for the full sample,
the fraction of galaxies with ``collision corrected'' redshifts is smaller 
than the case we examined here, and thus we expect the effects of the fiber 
collisions to be even smaller.

A detailed examination of fiber collision effects will require tests
on artificial catalogs with realistic galaxy clustering and geometry,
but the agreement between the dashed line and unity in 
Figure~\ref{fig:fiberc} implies that any residual systematic biases 
in our correlation function estimates due to fiber collisions are smaller 
than our current statistical errors.

\subsection{Angular Selection Function}
\label{subsec:ang_sel}

A small fraction of the galaxy targets in our sample were not assigned 
fibers in the observed plates. There are also some galaxy targets whose 
redshifts are not successfully measured, for the most part because of 
broken fibers in the spectrograph, but sometimes because of a low 
signal-to-noise ratio in the spectra. 
The completeness of the redshift sample, denoted here $f$, thus varies 
across the survey, and it is necessary to incorporate 
these variations into the window function of the survey.
We evaluate the completeness in the following way. We break up
the survey geometry into ``sectors'' defined by areas of sky covered 
by unique sets of tiles, as described by \citet{blanton01b}. For example, 
if the survey consisted of two tiles, there would be three sectors: the 
area covered by only the first tile, the area covered by only the second 
tile, and the overlap area covered by both tiles. 
These sectors are the natural units in which to divide the survey,
and we calculate the completeness $f$ for each sector. The completeness 
is simply the fraction of objects that were selected as galaxy targets for 
which a spectral classification was obtained (whether the object turned out 
to be a galaxy or not), or for which a redshift was assigned because the 
object was lost in a collision. 

In this sample,  the average completeness is about 94\%. 
There are two contributions to the incompleteness.  
First, only 97\% percent of the available galaxy targets in the regions 
covered by plates actually are assigned fibers or have a close neighbor 
that can provide a ``collision corrected'' redshift. This is partly 
because some galaxies are eliminated due to fiber collisions with QSOs, 
LRGs or other objects which have higher priority when fibers are assigned, 
and thus cannot be given collision corrections. In addition, we have 
included some regions that are covered by two plates, only one of which 
has so far been observed; the targets in such a region that are assigned 
to fibers on the unobserved plate contribute to the incompleteness.  
Second, the fraction of fibers assigned to main galaxy targets that 
successfully receive classifications and redshifts is about 97\% in this 
sample. The success rate for obtaining main sample galaxy 
redshifts during normal survey operations 
is over $99\%$. However, some of the data in this sample come from plates 
that have low signal-to-noise ratio (and will therefore be reobserved later 
in the survey) or were reduced using older, less efficient versions of the 
spectroscopic pipeline. In addition, 
some of these targets are imaging defects that were mistakenly classified as 
galaxies by early versions of the galaxy target selection algorithm, such 
as ghost images due to reflections of bright stars inside the camera or
satellite trails. Though these latter cases, in fact, do not contribute 
to galaxy incompleteness, they are included in our estimate of $f$, but 
this makes a negligible difference to our results.

We apply several masks for regions of particularly bad seeing and
where an early version of the tiling algorithm (now replaced)
accidentally produced artificial gaps in the sampling. We exclude any
objects in our data or random catalog that lie inside these
masks. The masks cover less than $1\%$ of the total area. We
have not applied masks around bright stars; if we did, they 
would exclude about $1\%$ of the total area (\citealt{scranton01}). It
will be necessary to include these masks when studying clustering at
the largest scales, because at large scales the clustering amplitude
of stars becomes large (due to the variation with Galactic latitude)
and the clustering amplitude of galaxies becomes small.

We properly take into account the incompleteness in each individual
sector when calculating the correlation function. But, in fact, 
because the completeness of the redshift sample is high to begin with,
the effects of completeness variations on our current clustering 
measurements are negligible. We have verified this by calculating the
correlation function, not accounting for the incompleteness, obtaining
almost indistinguishable results.

\subsection{Radial Selection Function}
\label{subsec:rad_sel}

As noted above, our sample is limited at bright and faint apparent
magnitudes: $14.5<\rs<17.6$.  Thus, at any given redshift we can
only observe galaxies in a given absolute magnitude
range. Furthermore, we restrict our analysis here to galaxies with
absolute magnitudes $-22<M_{\rs}<-19$. At any redshift, the fraction
of objects in this absolute magnitude range that are in the sample is
\begin{equation}
\phi(z) = 
\frac{\int_{M_{\mathrm{min}}(z)}^{M_{\mathrm{max}}(z)} dM \Phi(M)}
{\int_{-22}^{-19} dM \Phi(M)},
\label{eqn:phi}
\end{equation}
where $\Phi(M)$ is the luminosity function (number density of objects
per unit magnitude) and
\begin{eqnarray}
M_{\mathrm{min}}(z) &=& \mathrm{max}(-22, 14.5-\mathrm{DM}(z)-K(z)), \cr
M_{\mathrm{max}}(z) &=& \mathrm{min}(-19, 17.6-\mathrm{DM}(z)-K(z)),
\label{eqn:mminmax}
\end{eqnarray}
and $\mathrm{DM}(z) = m - M$ is the distance modulus as described in
Section \ref{subsec:cosmo}.  In this context, $K(z)$ is determined
using the mean galaxy SED in the sample.
Equations~(\ref{eqn:phi}) and~(\ref{eqn:mminmax}) simply express the
fact that a galaxy must lie in our apparent magnitude range and in
our absolute magnitude range to be included in the sample.

The luminosity function for our sample is determined in the manner
described by \citet{blanton01a}.  It is necessary to perform this
calculation separately for each of the subsamples described in Section
\ref{sec:dependencies} because the luminosity functions of, for
example, blue galaxies and red galaxies differ substantially.  The
luminosity function for our full sample is consistent with that of
\citet{blanton01a} when determined using the same redshift limits as
that paper. However, we note here that it appears from preliminary
results (to be described in detail elsewhere) that the galaxy
luminosity function evolves measurably within the redshift range of our
spectroscopic sample. At the time the calculations presented here were 
performed, we had not yet accounted for this effect in our calculation 
of the selection function. This is our main motivation for limiting the 
current sample to $cz < 39,000\kms$. More recently, we have fit a pure 
luminosity evolution model to the data.
The resulting change in the selection function below $cz = 39,000\kms$ 
is less than $5\%$, and the resulting differences in the measured 
correlation functions are negligible. Thus we are confident that our 
radial selection function calculated without accounting for evolution 
is sufficient to study the small-scale clustering of interest here.

\begin{figure}[tbp]
{\includegraphics{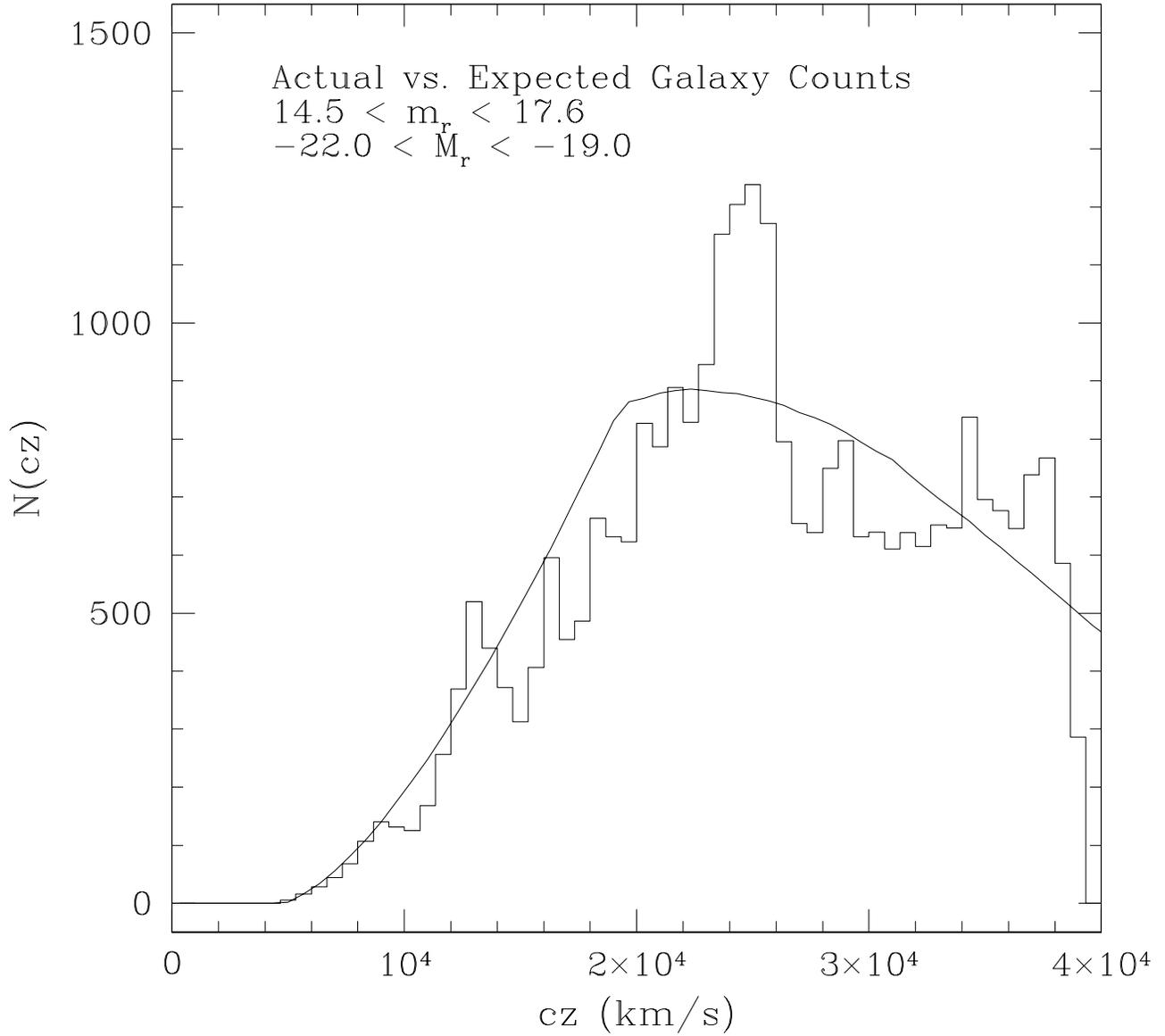}}
\vspace{16 cm}
\caption{Histogram of the redshift distribution of the SDSS galaxies in 
our sample. The solid line is the average distribution expected given
the luminosity function, the flux limits, and the angular selection
function.}
\label{fig:dndz}
\end{figure}

When the random sample is created for the calculation of the correlation 
function (see below), this selection function $\phi(z)$ and the local
completeness $f$ must be taken into account. In practice, we first
distribute points uniformly in comoving space; we then include each
such point in the random sample with a probability $f \phi(z)$. 
Figure \ref{fig:dndz} compares the expected redshift
distribution of this uniform sample (smooth line) to the actual
redshift histogram of galaxies, including the galaxies whose redshifts
were assigned by our collision correction method. The differences between 
the expected redshift distribution and the actual one reflect the 
large-scale structure which we are here attempting to measure.

\subsection{Estimator}
\label{subsec:estimate}

To account for the survey geometry, we generate random catalogs 
of galaxies with the same survey geometry as the real sample,
applying both the radial and angular selection functions. 
We typically use in each random catalog 10 times the number of 
galaxies in the real sample, and we have verified that further increasing
the number of random points makes negligible difference to the results.

We calculate the correlation function using the \citet{landy93}
estimator, 
\begin{equation}
\xi(s)= \frac{1}{N_{RR}(s)}
\left[N_{DD}(s)\left(\frac{n_R}{n_D}\right)^2-2N_{DR}(s)
 \frac{n_R}{n_D}+N_{RR}(s)\right],
\label{eq:LS}
\end{equation}
where $N_{DD}$, $N_{DR}$, and $N_{RR}$ are the weighted data-data, 
data-random, and random-random pair counts, respectively,  with 
redshift-space separations in a bin centered on $s$, and $n_D$ and 
$n_R$ are the mean number densities of galaxies in the data and random 
samples. Bins in $s$ are logarithmically spaced with width of 0.2 in 
$\log(s/\hmpc)$ starting from $s=0.1\hmpc$. 
Other statistics are calculated in an analogous way.
We also tried the alternative estimators of \citet{davis83} and
\citet{hamilton93} and found no significant difference in the results.

For the pair weighting we follow \citet{hamilton93} and use a minimum
variance weighting (see also \citealt{davis82}; \citealt{feldman94}).
For a galaxy pair with redshift separation $s$, we weight each galaxy by
\begin{equation}
w_i = \frac{1}{1+4\pi n_D f_i\phi(z_i)J_3(s)}\,,               
\label{eq:MV}
\end{equation}
where $J_3(s) \equiv \int_0^{s} s'^2 \xi(s') ds'$. For this integral, we
approximate $\xi$ by a power-law with slope $-1.2$ and correlation
length $8\hmpc$ (resembling the result for the correlation function in
redshift space that we get below, see \S~\ref{subsec:sspace}), but the
results are robust to reasonable choices. Alternatively, we also
weighted each galaxy simply by the inverse of the (radial and angular) 
selection function and obtained comparable results.

The full covariance error matrices are obtained by a jackknife error
estimate (see, \eg, \citealt{lupton93}). We divide our sample into ten
separate regions on the sky of approximately equal area. We perform the 
analysis ten times, each time leaving a different region out. The 
estimated statistical covariance of $\xi_i$ in redshift separation 
bin $i$ and  $\xi_j$ in bin $j$ is then
\begin{equation}
\label{eq:jk}
{\rm Covar}(\xi_i,\xi_j) = \frac{N-1}{N} \sum_{l=1}^{N} 
({\xi_i}^l - {\bar{\xi}_i}^l)({\xi_j}^l - {\bar{\xi}_j}^l),
\end{equation}
where $N=10$ in our case, and $\bar{\xi}_i$ is the mean value of 
$\xi_i$ measured
in the samples. Further discussion regarding the robustness of the
jackknife error estimate and comparison to alternative error estimates
can be found in Appendix A.
Note that if the number of regions is increased ($N>10$), then each 
term in the sum decreases (because the $N-1$ regions in each jackknife
subsample are a larger fraction of the total sample), but the number
of terms increases, so the estimated covariance converges to a stable
answer.

In what follows, we present results for the Landy-Szalay estimator, 
with minimum variance weighting for the galaxies, and errors obtained
by jackknife resampling. 
Galaxies with missing redshifts due to fiber collisions are accounted 
for as described above in Section \ref{subsec:fiberc}.

\section{Clustering of the Full Sample}
\label{sec:results}

In this Section we present the results for our full galaxy sample.
Summarizing the details described in the previous sections, the 
sample consists of $29,300$ galaxies with
redshift $5,700 \kms < cz < 39,000 \kms$, 
apparent magnitude (corrected for Galactic absorption) $14.5<\rs<17.6$, 
and absolute magnitude $-22<M_{\rs}<-19$. 

\subsection{Redshift-Space Clustering}
\label{subsec:sspace}

\begin{figure}[tbp]
{\includegraphics{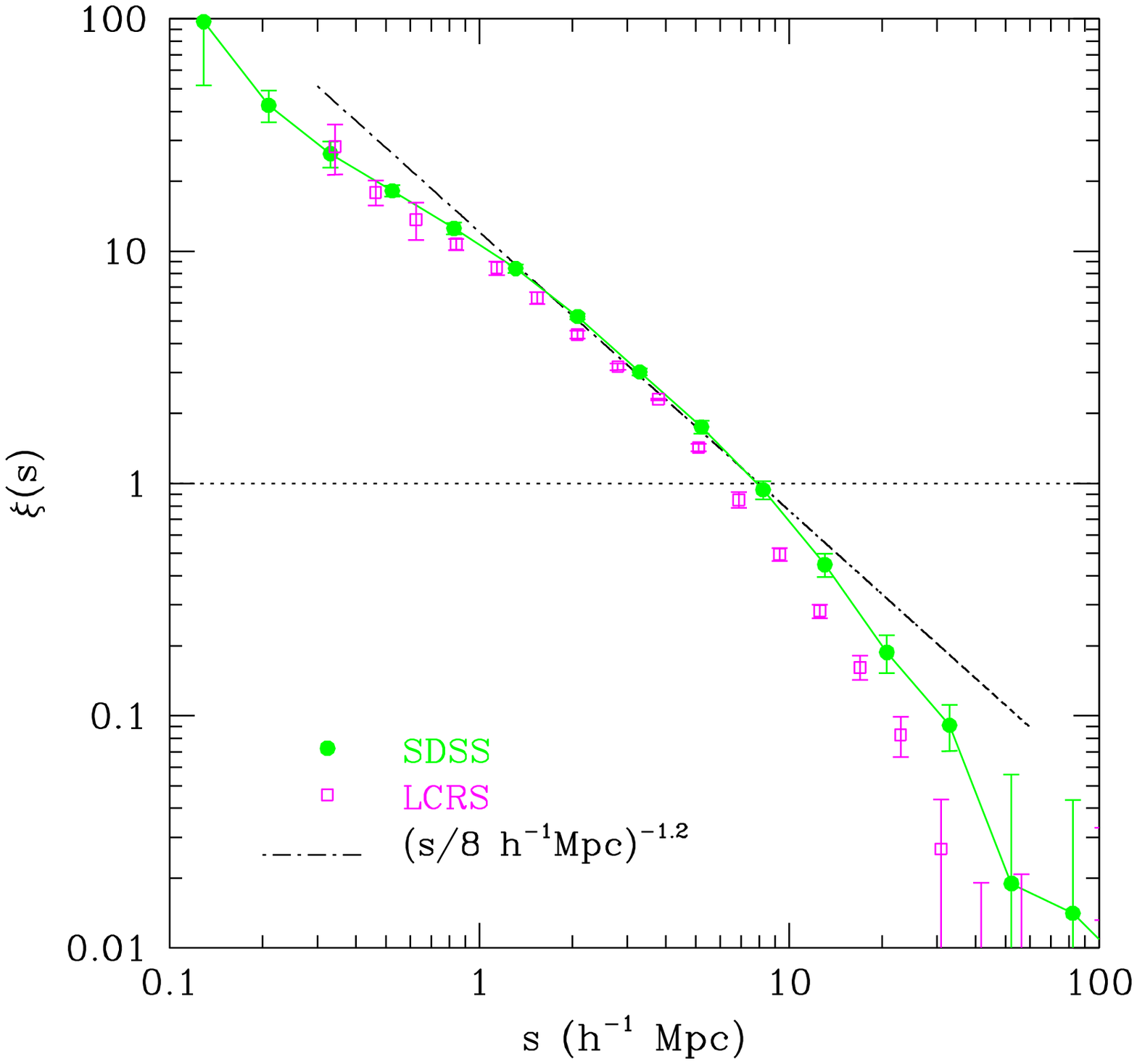}}
\vspace{16 cm}
\caption{The redshift-space correlation function $\xi(s)$ (solid points
and line). The error bars plotted here, and in all subsequent figures, 
correspond to the $1 \sigma$ uncertainty estimated from jackknife resampling. 
A fiducial power-law fit 
for the range $2 \hmpc < s < 10 \hmpc$ is plotted as a dot-dashed line.
Open squares show $\xi(s)$ obtained from the LCRS (\citealt{tucker97}).
} 
\label{fig:xsis}
\end{figure}

Figure~\ref{fig:xsis} shows the redshift-space correlation function
$\xi(s)$ of the full sample. For separations 
$2\hmpc < s < 10\hmpc$, the observed correlation function can be 
crudely approximated by a power-law, $\xi(s)=(s/s_0)^{-\gamma}$, 
with $\gamma_s=1.2$ and $s_0=8.0 \hmpc$.
Table~\ref{table:comp} summarizes our results for the full sample
together with the corresponding results obtained for some other major
redshift surveys available in the literature.
In our comparison to other surveys, we focus largely on the LCRS 
\citep{shectman96}, as this survey resembles ours most closely in terms
of selection, geometry, and analysis. Their results for the redshift-space 
correlation function are shown as well in Figure~\ref{fig:xsis} (open 
squares; taken from \citealt{tucker97}). The LCRS results are in quite
good agreement with ours,
though the SDSS correlation function has a slightly higher amplitude.
We have assumed an $\Omega_m=0.3$, $\Omega_\Lambda=0.7$ model to
compute comoving separations, but adopting an Einstein-de Sitter
model (EdS; as Tucker et al.\ do) yields a nearly indistinguishable result.
The SDSS $\xi(s)$ remains measurably non-zero out to $s=30\hmpc$ 
and is consistent with zero at larger separations.

\begin{deluxetable}{ccccccc}
\tablewidth{0pt}
\tablecolumns{7}
\tablecaption{\label{table:comp} Clustering Results of Different
Galaxy Redshift Surveys}
\tablehead{ Survey & $N_{\mathrm{gal}}$ & $s_0$ & $\gamma_s$ & $r_0$ & 
$\gamma$ & $\sigma_{12}(1\hmpc)$}
\tablecomments{$s_0$ and $r_0$ are in units of $\hmpc$, $\sigma_{12}$ is in
units of $\kms$.\\
$^a$ We use comoving distances assuming $\Omega_m=0.3$
$\Omega_\Lambda=0.7$. With an Einstein de-Sitter model we get $r_0=5.7 \pm
0.2$ and $\sigma_{12}(1\hmpc)=590 \pm 50$. Note that a power-law is a 
poor fit to $\xi(s)$, though a good fit to $\xi_r(r)$. \\
$^b$ \citealt{norberg01}; these are the fit parameters for a 
volume-limited sample of galaxies with $-19.5 < M_{b_J} < -20$,
close to $M^* = -19.7$ \citep{folkes99}. \\
$^c$ Here 15123 refers to a volume-limited sample, drawn from a 
flux-limited sample containing $\sim 160,000$ galaxies. \\
$^d$ \citealt{tucker97,jing98} (both assuming an EdS model). \\ 
$^e$ \citealt{jing01}, using  9,400 galaxies (EdS cosmology).
As galaxies are selected from
the IRAS catalog, they are preferentially late types, and thus are
more directly comparable to our ``blue'' galaxies sample, see
\S\ref{subsec:color}. \\
$^f$ $s_0$ and $\gamma_s$ taken from \citealt{delapparent88}, 
using 1,800 galaxies of first slice; $r_0$ and $\gamma$ based on Fig.~3 
of Marzke et al.'s (1995) analysis of CfA2 and SSRS2; $\sigma_{12}$ from 
\citealt{marzke95}. \\
$^g$ \citealt{hermit96}. 
}
\startdata
SDSS$^a$ & 29,300 & $\sim$8.0 & $\sim$1.2 & 6.14 $\pm$ 0.18 & 1.75 $\pm$ 0.03 & 640 $\pm$ 60 \cr 
2dF$^b$ & 15,123$^c$ & -- & -- & 4.92 $\pm$ 0.27 & 1.71 $\pm$ 0.06 & -- \cr
LCRS$^d$ & 26,400 & 6.3 $\pm$ 0.3 & 1.52 $\pm$ 0.03 & 5.06 $\pm$ 0.12 & 1.86 $\pm$ 0.03  & 570 $\pm$ 80 \cr
PSCz$^e$ & 15,400 & 5.0 & 1.2 & 3.7 & 1.69 & 350 $\pm$ 60 \cr
CfA2$^f$ & 12,800 & $\sim$7.5 & $\sim$1.6  & 5.8 & 1.8 & 540 $\pm$ 180 \cr 
ORS$^g$ & 8,500 & 7.6 $\pm$ 1.2 & 1.6 $\pm$ 0.1 & 6.1 $\pm$ 1.2 & 1.6 $\pm$ 0.1 & -- \cr 
\enddata
\end{deluxetable}

The redshift-space correlation function $\xi(s)$ differs from the 
real-space correlation function $\xi_r(r)$ because of peculiar velocities.
Following standard practice,
we separate the  effects of redshift-space distortions from spatial
correlations by separating the vector between two galaxies into a 
line-of-sight component $\pi$ and a projected component $r_p$,
and measuring $\xi(r_p,\pi)$.
More specifically, following the notation of \citet{fisher94}, for a pair of 
galaxies with redshift positions ${\bf v}_1$
and ${\bf v}_2$, we define the redshift separation vector
${\bf s} \equiv {\bf v}_1-{\bf v}_2$ and the line-of-sight vector
${\bf l} \equiv {{1}\over{2}}({\bf v}_1+{\bf v}_2)$. This allows us to 
define the parallel and perpendicular separations
\begin{equation}
\label{eq:rppi}
\pi \equiv {\bf s}\cdot{\bf l}/|{\bf l}|\,, \qquad
{r_p}^2 \equiv {\bf s}\cdot{\bf s} - \pi^2\,.
\end{equation}
In real space, the contours of equal $\xi$ should be circular (by isotropy, 
$\xi$ depends only on the scalar separation), but in redshift space the 
contours are distorted by peculiar velocities.

\begin{figure}[tbp]
{\includegraphics{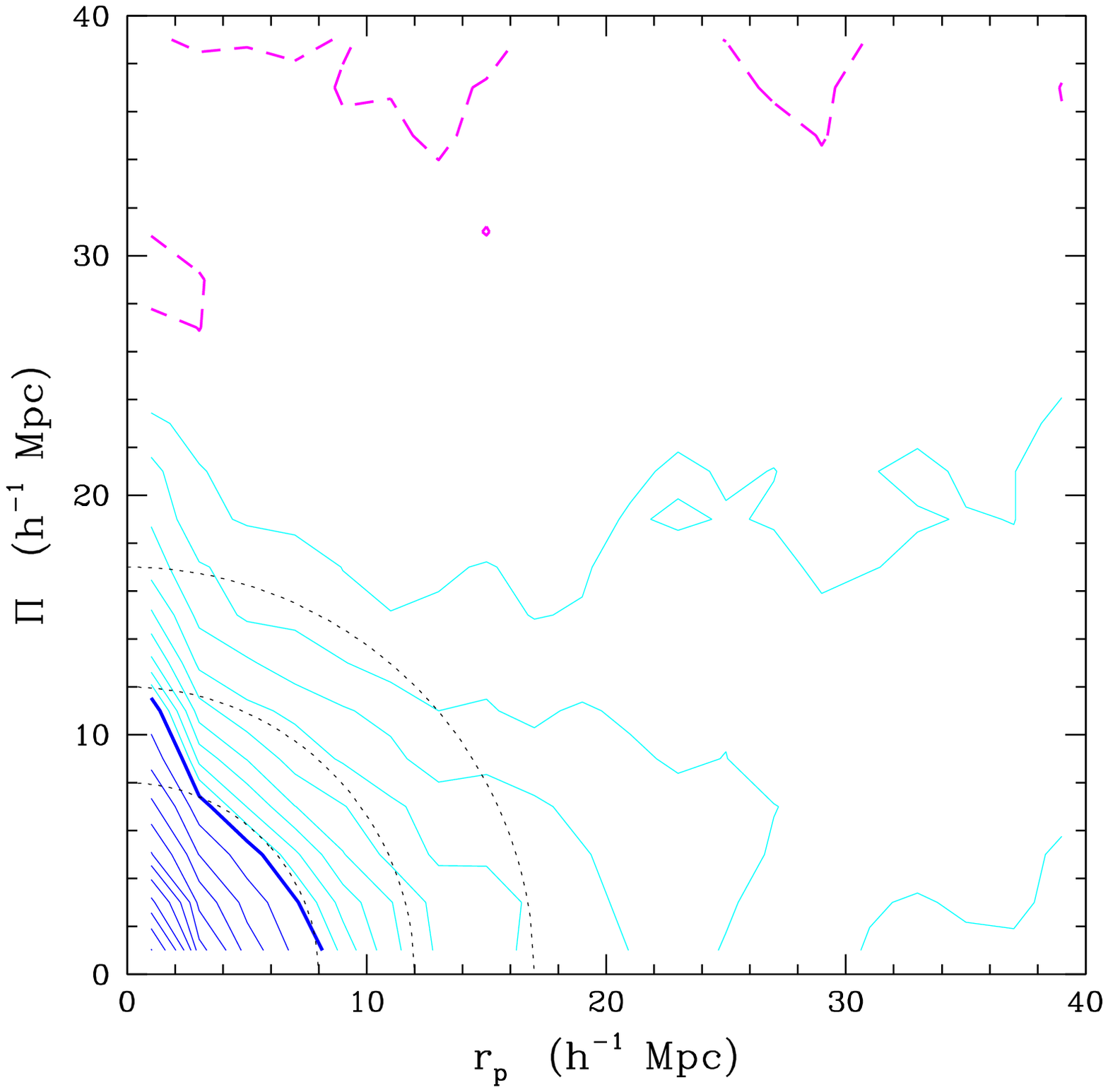}}
\vspace{16 cm}
\caption{Contours of $\xi(r_p,\pi)$, the correlation function as a function of 
separation perpendicular ($r_p$) and parallel ($\pi$) to the line of sight.
The heavy solid contour corresponds to $\xi=1$; for larger values of
$\xi$ contours are logarithmically spaced, with $\Delta \log_{10}\xi = 0.1$;
below $\xi=1$ they are linearly spaced, with $\Delta \xi = 0.1$;
the heavy dashed contour corresponds to $\xi=0$. The concentric dotted 
lines are the angle-averaged redshift-space correlation
function, $\xi(s)$, at $\xi(s) = 1.0$, $0.5$ and $0.25$.}
\label{fig:xsirpi}
\end{figure}

Figure~\ref{fig:xsirpi} shows $\xi(r_p,\pi)$ for our sample, where we
bin $r_p$ and $\pi$ in linear bins of $2\hmpc$. 
On small scales, the contours are elongated along the line of sight 
direction, exhibiting the expected ``fingers-of-God'' distortion caused by
velocity dispersion in collapsed objects. On larger scales, $\xi(r_p,\pi)$
shows compression in the $\pi$ direction, caused by coherent 
large-scale streaming.  The qualitative appearance of Figure~\ref{fig:xsirpi}
is similar to that of, \eg, figure~1 of \citet{fisher94} 
or figure~2 of \citet{peacock01}.

\subsection{Real-Space Clustering}
\label{subsec:rspace}

The effects of redshift-space distortions are only radial, so projection 
onto the $r_p$ axis gives information about the real-space correlation
function.   We compute the projected correlation function $w_p(r_p)$
by integrating $\xi(r_p,\pi)$ over $\pi$,
\begin{equation}
w_p(r_p) \equiv 2 \int_0^{\infty} d\pi \, \xi(r_p,\pi) = 2 \int_0^{\infty}
dy \, \xi_r\left(\sqrt{r_p^2+y^2}\right),
\label{eq:wp}
\end{equation}
where $\xi_r$ is the desired real-space correlation function \citep{davis83}.
In practice we integrate up to $\pi_{max}=40 \hmpc$, which is large
enough to include most correlated pairs and to give a stable result.
The second equation (rhs) above allows us to relate $w_p$ to the real-space
correlation function. In particular, for a power-law 
$\xi_r(r) = (r/r_0)^{-\gamma}$,
the second integral can be done analytically, yielding 
\begin{equation}
w_p(r_p)=A r_p^{1-\gamma}\, \quad {\rm with} \quad
A=r_0^{\gamma} \Gamma(0.5) \Gamma[0.5(\gamma-1)]/\Gamma(0.5\gamma),
\label{eq:wp2}
\end{equation}
where $\Gamma$ is the Gamma function.

\begin{figure}[tbp]
{\includegraphics{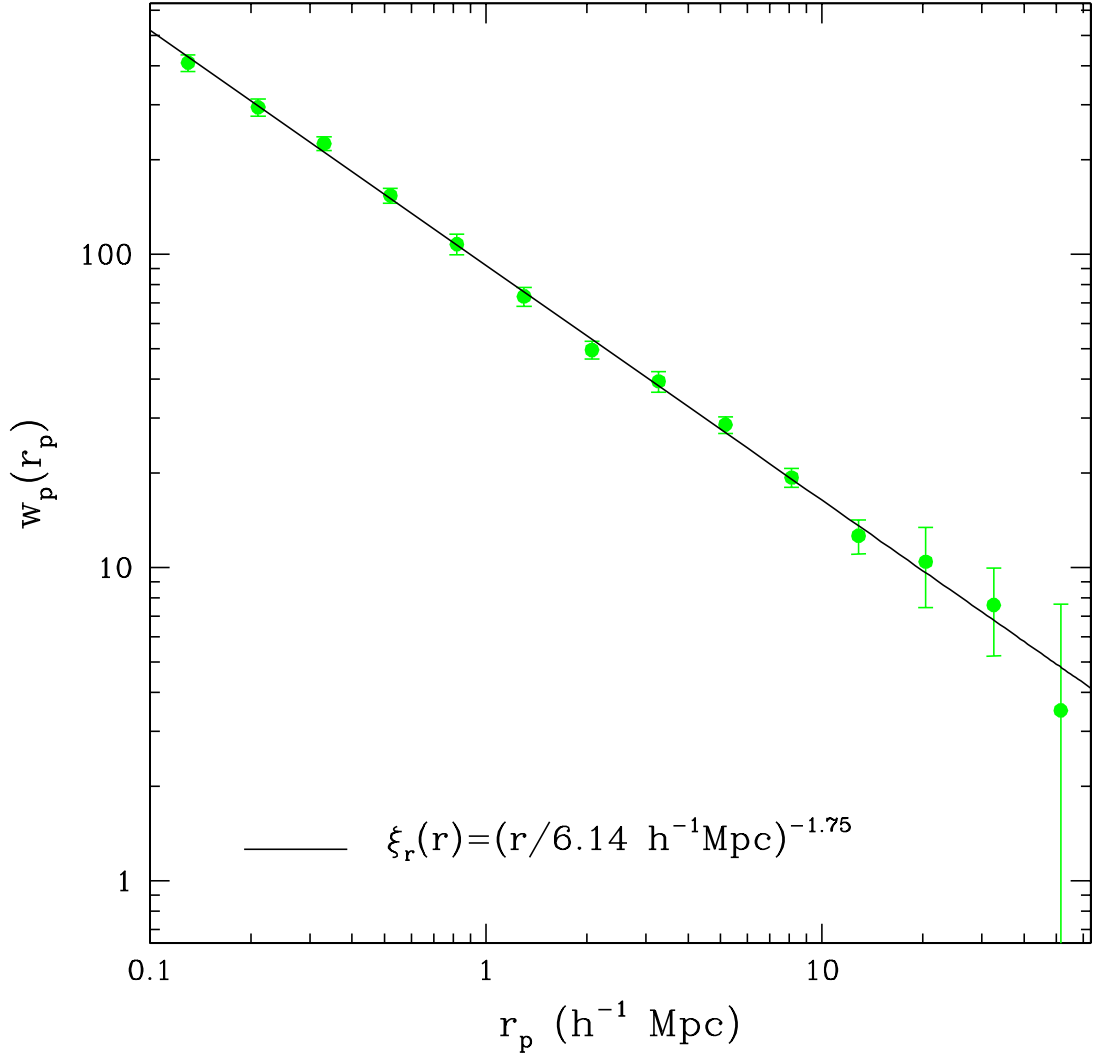}}
\vspace{16 cm}
\caption{Projected correlation function $w_p(r_p)$ (solid points). The 
solid line is the best-fit  power-law for $w_p$, which implies the denoted 
power-law for the real-space correlation function $\xi_r(r)$. The fit is
performed for $r_p < 16\hmpc$. }
\label{fig:wp}
\end{figure}

Figure~\ref{fig:wp} shows $w_p(r_p)$ for the full galaxy sample
and the best-fit power-law model, which corresponds to 
$\xi_r(r)=(r/r_0)^{-\gamma}$ with $r_0=6.14 \pm0.18 \hmpc$, 
$\gamma=1.75 \pm0.03$.  This fit to the slope and amplitude of 
the correlation function is obtained using points
in the range $0.1\hmpc < r_p < 16\hmpc$;   
the correlation coefficient between $r_0$ and $\gamma$, measuring the
normalized covariance of the two estimates, is $\sim -0.5$, implying that 
the measures are anti-correlated to a degree. 
Since the jackknife
estimates of the off-diagonal terms in the covariance matrix are noisy  
and lead to an unstable matrix inversion in the $\chi^2$ minimization 
(unless we confine the fit to only a few bins), 
the best-fit $r_0$ and $\gamma$ values were obtained 
from the diagonal terms only. As a result, 
we are not guaranteed to have unbiased estimates of these parameters, but 
the visually evident goodness-of-fit suggests that any such bias is negligible.
The errors on $r_0$ and $\gamma$  were obtained from the 
variance in the estimates of these quantities among the 
jackknife subsamples, again using only the diagonal terms in the 
covariance matrix, as described in Appendix A. 

The real-space correlation function is characterized much more 
accurately by a power-law than the redshift-space correlation function.  
Our value of $\gamma$ agrees well with results from previous redshift 
surveys and angular clustering studies 
(\eg, \citealt{davis83}; \citealt{loveday95}; Table~\ref{table:comp}) 
and with the slope derived from the SDSS angular correlation
function \citep{connolly01}.  The value of $r_0$ is also similar to 
that obtained from other optically selected galaxy samples, 
as can be seen in Table~\ref{table:comp}, though in some cases 
slightly on the high side; for example, Jing, Mo, \& B\"orner (1998)
find $r_0=5.06\pm 0.12\hmpc$ for the LCRS.  If we adopt an EdS 
cosmology, as they do,  instead of a flat-$\Lambda$ cosmology, 
then our inferred value of $r_0$ drops slightly, to $r_0=5.7\pm 0.2\hmpc$.
Remaining differences may be attributed to the specifics of the
selection criteria and magnitude ranges, reflecting the dependence of
galaxy clustering on color and luminosity.
Our value of $r_0$ is also higher than the value $r_0=4.92\pm 0.27$ 
found for $b_J$-selected galaxies with $M \sim M_*$ in the
2dFGRS by Norberg et al.\ (2001), but it is similar to the value
they obtain for galaxies 0.5-1 magnitudes brighter than $M_*$
(see their table 1).

\subsection{Angular Moments}
\label{subsec:moments}

The redshift-space correlation function $\xi(s)$ in Figure~\ref{fig:xsis}
differs from the real-space correlation function $\xi_r(r)$ inferred
from $w_p(r_p)$ in the expected sense: $\xi(s)$ is depressed on small
scales by velocity dispersions and enhanced on large scales by coherent
flows, so the slope of $\xi(s)$ is shallower and $s_0>r_0$.
The anisotropy of $\xi(r_p,\pi)$ encodes more complete information
about the amplitude of galaxy peculiar velocities.
In principle, the anisotropy on large scales can be used to 
constrain $\beta \equiv \Omega_m^{0.6}/b$, where the bias parameter $b$
is the ratio of galaxy fluctuations to mass fluctuations
\citep{kaiser87,hamilton92}.  For this application, it is helpful to 
decompose $\xi(r_p,\pi)$ into a sum of Legendre polynomials,  
\begin{equation}
\xi(r_p,\pi)=\sum_l \xi_l(s){\cal P}_l(\mu),
\end{equation}
where ${\cal P}_l$ is the $l^{th}$ Legendre polynomial and $\mu$ is the cosine
of the angle between the line of sight and the redshift separation vector.
The angular moments are found by integration
\begin{equation}
\xi_l(s)=\frac{2l+1}{2} \int_{-1}^{1} \xi(r_p,\pi) {\cal P}_l(\mu)d\mu.
\end{equation}
In linear perturbation theory, only the monopole, $\xi_0(s)$, quadrupole,
$\xi_2(s)$, and hexadecapole, $\xi_4(s)$ are non-zero, and the ratio
\begin{equation}
\label{eq:G}
Q(s) \equiv \frac{\xi_2(s)}{\frac{3}{s^2}\int_{0}^{s} 
\xi_0(s')s'^2ds' -\xi_0(s)}  = 
G(\beta) \equiv
\frac{\frac{4}{3}\beta+\frac{4}{7}\beta^2}{1+\frac{2}{3}\beta+
\frac{1}{5}\beta^2}
\end{equation}
\citep{hamilton92}.  Thus, the ratio $Q$ provides an estimate of $\beta$
(similar estimates can be constructed using $\xi_4(s)$, but they are 
noisier).  However, while linear theory distortions produce a negative
quadrupole term, finger-of-God distortions produce a positive quadrupole,
and their signature persists out to large separations,
(\citealt{cole94,fisher94}). 

\begin{figure}[tbp]
{\includegraphics{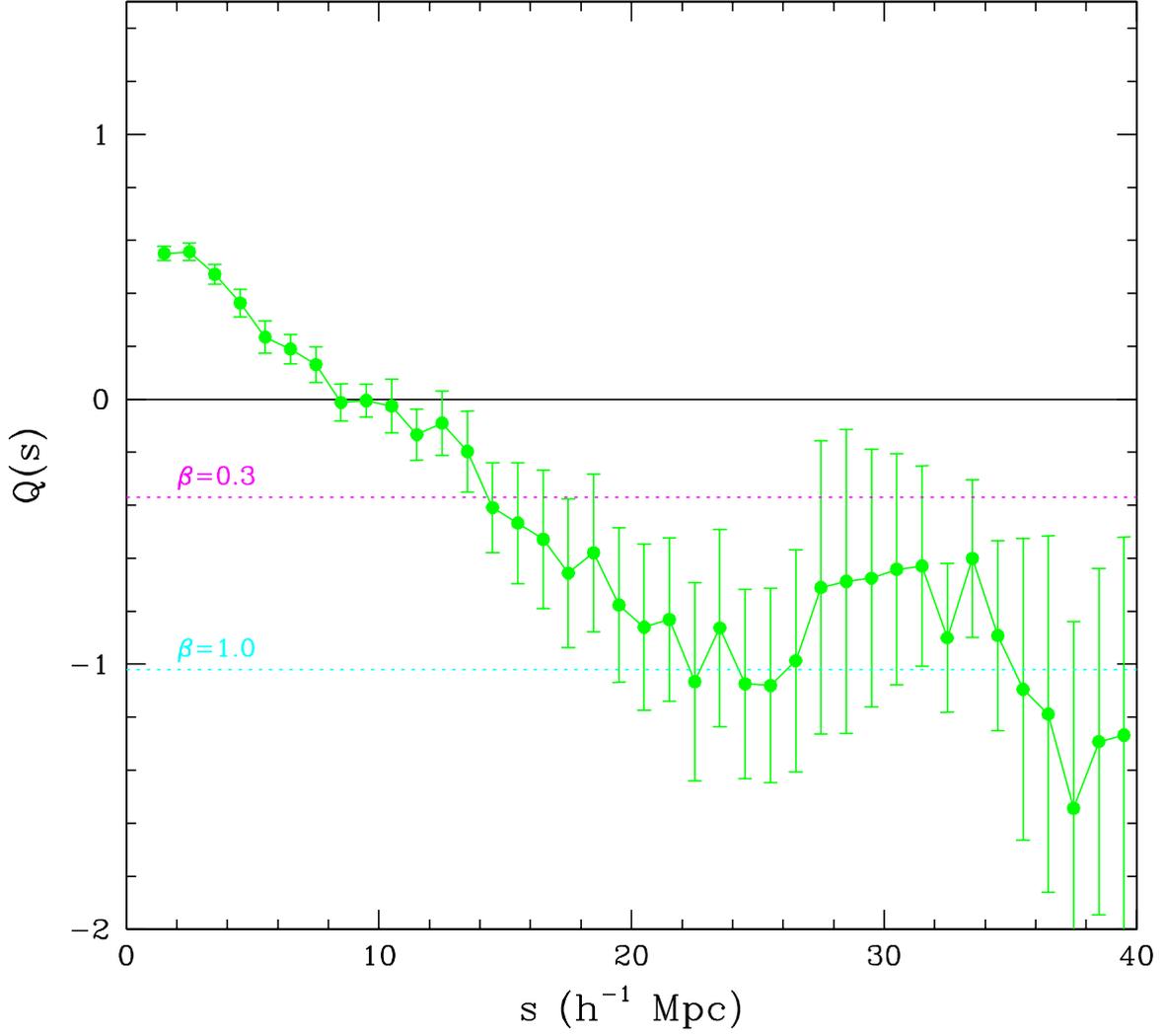}}
\vspace{16 cm}
\caption{Modified quadrupole to monopole ratio, $Q=\xi_2/(\bar{\xi_0}-\xi_0)$.
In linear theory this ratio is determined by the parameter
$\beta \equiv \Omega_m^{0.6}/b$. Dotted lines show the linear theory 
expectation for $\beta=0.3$ and $\beta=1.0$. }
\label{fig:qlin}
\end{figure}

Figure~\ref{fig:qlin} shows the quadrupole ratio $Q(s)$ for the full
sample. The error bars are obtained, as before, from the scatter in
the jackknife subsamples.  This figure
quantifies the visual impression of the contours in
Figure~\ref{fig:xsirpi}, showing positive (finger-of-God) quadrupole 
distortion at $s \la 10\hmpc$ and negative (coherent flow) quadrupole
distortion at larger scales.  $Q(s)$ should approach a constant value
in the linear regime, and the measured results are consistent with
this prediction.  However, the error bars on these scales are large and
highly correlated, whereas high precision over a range of scales is
needed to separate the influence of coherent flows from that of
small-scale dispersions (see, \eg, \citealt{hatton98}).
The effective volume of our current sample is $\sim 4 \times 10^6 (\hmpc)^3$. 
Our measurement of large-scale redshift-space distortions is
limited by finite volume effects, as a small number of
elongated superclusters and filaments in the data can give rise to
anisotropy in $\xi(r_p,\pi)$ on large scales. 
In this respect, the thin-slice geometry of our present sample works
against us, since it provides relatively few pairs at large transverse
separations.  We therefore defer an estimate of $\beta$ to a future
study based on a larger, more nearly 3-dimensional sample of SDSS data,
and focus instead on the amplitude of small-scale, incoherent velocities
(but see \citealt{peacock01} for an estimate of $\beta$ from the 2dFGRS 
survey using a similar statistic). 

\subsection{Pairwise Velocity Dispersion}
\label{subsec:pvd}

In the non-linear regime, where density and velocity fields are weakly
coupled, the correlation function $\xi(r_p,\pi)$ can be modeled as 
a convolution of $\xi_r(r)$ with the galaxy pairwise velocity distribution
$F(V)$ (\citealt{peebles80}, \S 76; \citealt{davis83}; see \citealt{fisher95a}
for an illuminating discussion of the assumptions implicit in this 
approach).  If $F$ varies only slowly with $r$, one can write
\begin{equation}
\label{eq:pwise1}
1+\xi(r_p,\pi)
=H_0\int_{-\infty}^\infty 
\left[1+\xi_r\left(\sqrt{r_p^2+y^2}\right)\right] F(V)\,dy\,,
\end{equation}
where
\begin{equation}
V \equiv \pi -H_0 y+{\overline v_{12}}(r)
\end{equation}
and ${\overline v_{12}}(r)$ is the mean radial pairwise velocity 
of galaxies at separation $r$.
The real-space correlation function $\xi_r(r)$ can be inferred from
$w_p(r_p)$ as described in \S\ref{subsec:rspace}.
Unfortunately, the forms of ${\overline v_{12}}(r)$ and $F(V)$ are not
known {\it a priori} for galaxies.
Following \citet{davis83}, we assume that ${\overline v_{12}}(r)$ in the 
above equation takes the form 
\begin{equation}
\label{eq:v12}
{\overline v_{12}}(r) ={H_0y\over 1+(r/r_0)^2}\,.
\end{equation}
This model is based on the similarity solution of the pair conservation 
equation \citep{davis77}. Using the formulae given in \citet{mo97}, it can 
be shown that equation~(\ref{eq:v12}) with $r_0\sim 5 \hmpc$ matches 
reasonably well the mean streaming velocities of dark matter particles 
in the $\Lambda$CDM model with $\Omega_m=0.3$ and $\sigma_8\sim 1$. The 
similarity solution may therefore be a reasonable approximation for the 
underlying density field over limited ranges of length and time scales. 
If galaxies are biased relative to the mass with a constant bias factor 
independent of time, the mean streaming velocities for galaxies should 
have a similar form  (see \citealt{fisher94}). Our following presentation 
is based on equation~(\ref{eq:v12}), but we will test the 
sensitivity of our results to this assumed infall model.

Based on observational (\citealt{davis83,fisher94,marzke95}) and 
theoretical (\eg, \citealt{peebles76,diaferio96,sheth96,juszkiewicz98})
considerations, we adopt an exponential form for $F$,
\begin{equation}
F(V)=C\exp\left(-2^{1/2}\vert V\vert/\sigma_{12}\right)\,,
\end{equation}
where $C$ is a normalization factor and $\sigma_{12}(r)$ is the pairwise 
velocity dispersion (PVD). Under these assumptions, we can estimate 
$\sigma_{12}(r)$ by performing a $\chi^2$ minimization of  the difference 
between the observed
$\xi(r_p, \pi)$ and the prediction given by  equation~(\ref{eq:pwise1}). 
In practice, we minimize the following quantity,
\begin{equation}
\sum_i \left[ \frac{\xi^{\rm obs}(r_p,\pi_i)-\xi^{\rm pred}
       (r_p,\pi_i;\sigma_{12})}
       {\sigma_{\xi}^{\rm obs}(r_p,\pi_i)} \right]^2 ,
\end{equation}
where the summation is done over $\pi$ bins up to $15 \hmpc$ for
a fixed $r_p$, so generally $\sigma_{12}$ is a function of $r_p$.
Here $\sigma_{\xi}^{\rm obs}(r_p,\pi)$ is the error on $\xi(r_p,\pi)$ 
estimated from the jackknife samples. The fit for $\sigma_{12}$ is 
robust to changing the limiting $\pi$ in the range $10-20\hmpc$.

\begin{figure}[tbp]
{\includegraphics{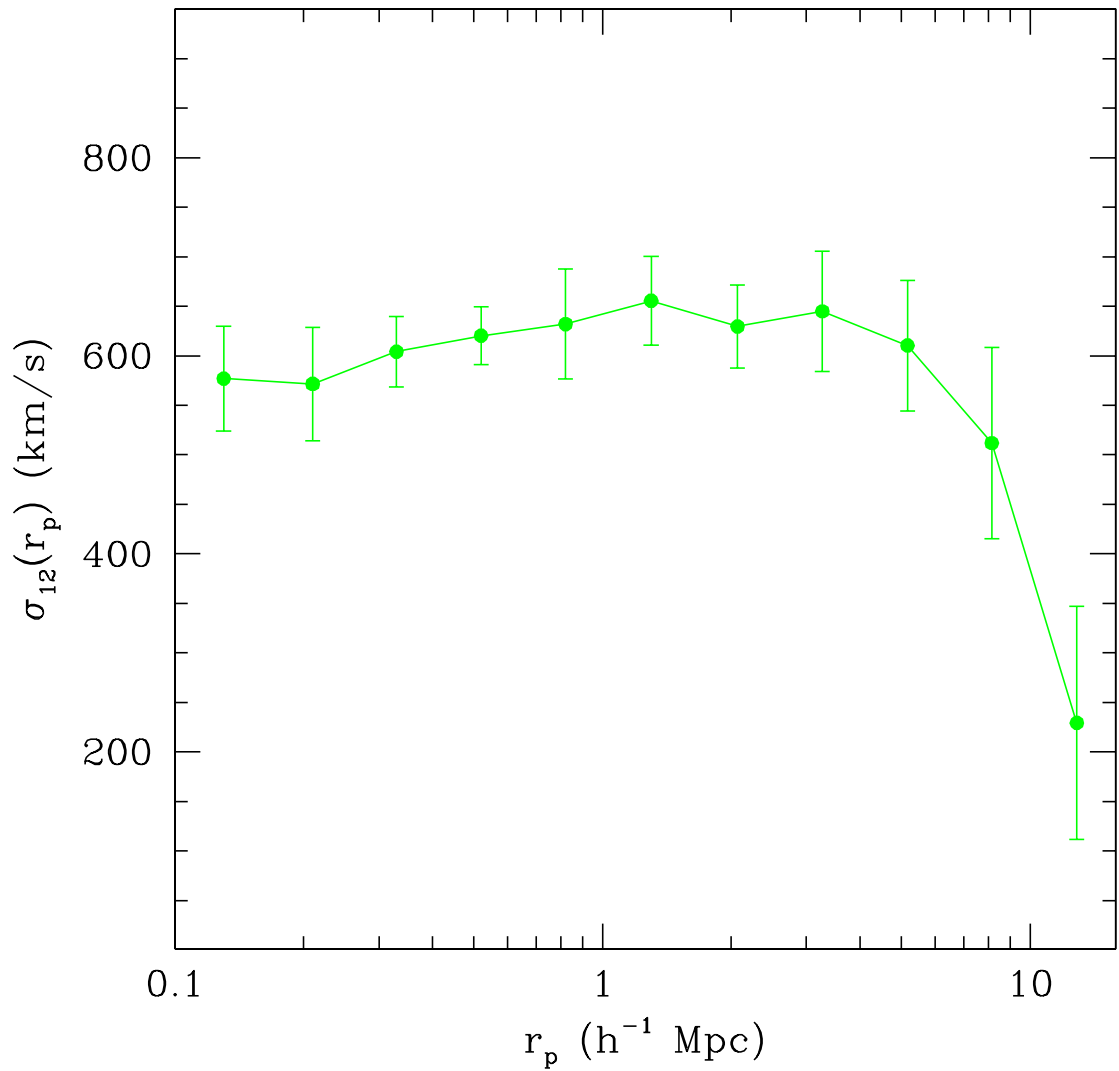}}
\vspace{16 cm}
\caption{The pairwise velocity dispersion $\sigma_{12}(r_p)$, inferred
by fitting $\xi(r_p,\pi)$.  Error bars are obtained from the values
of $\sigma_{12}$ in different jackknife subsamples.
The value of $\sigma_{12}$ at $r_p>3\hmpc$ depends significantly
on the assumed mean streaming model.
}
\label{fig:pvd}
\end{figure}

\begin{figure}[tbp]
{\includegraphics{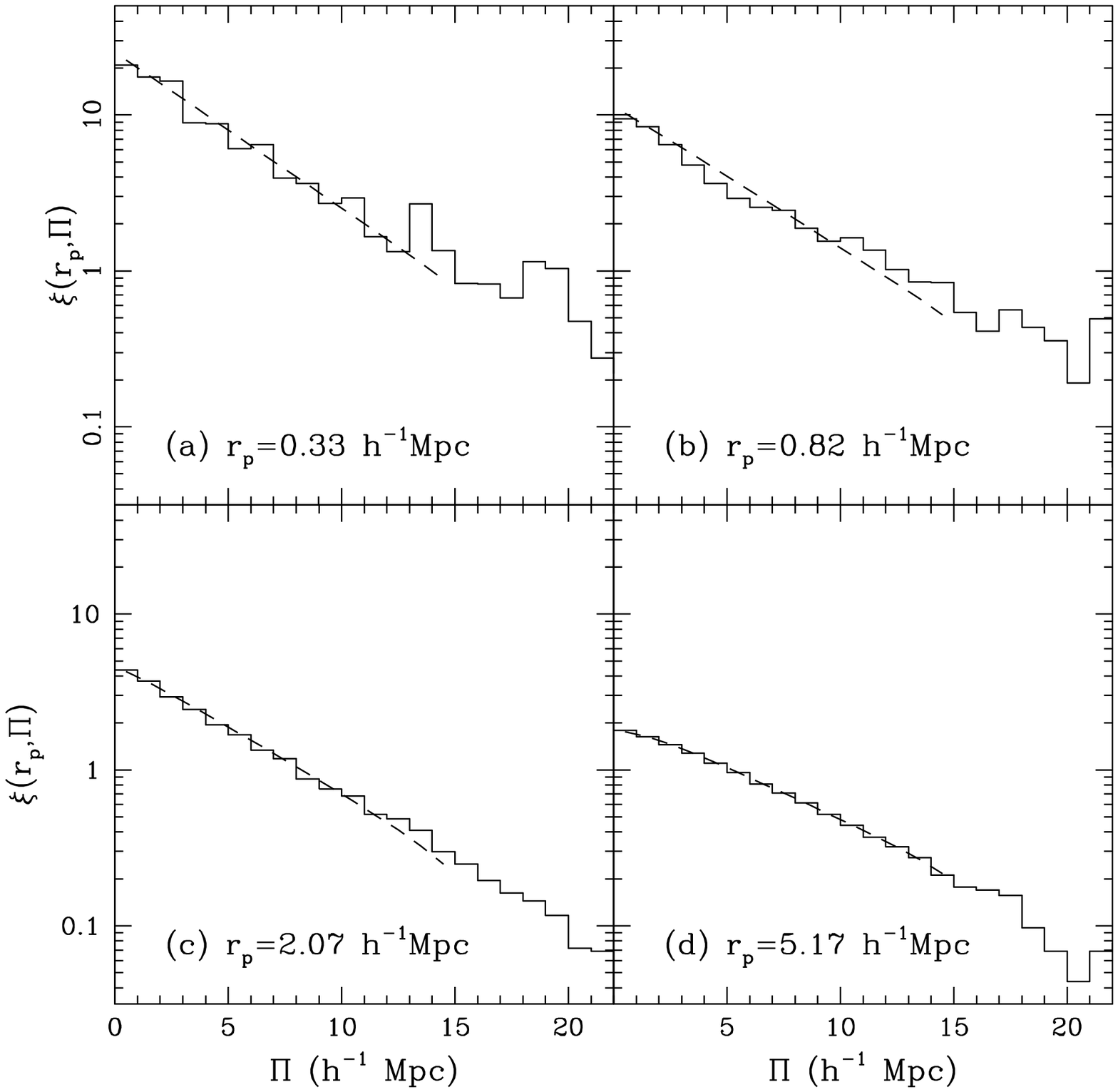}}
\vspace{17 cm}
\caption{Examples of the model fits for $\xi(r_p,\pi)$ for four
different values of $r_p$. The histogram shows the observed
values, and the dashed line is the model fit of eq.~(\ref{eq:pwise1}). }
\label{fig:xi_pi}
\end{figure}

Figure~\ref{fig:pvd} shows the result of this calculation, the PVD of the
full sample for projected separations $0.1\hmpc<r_p<20\hmpc$.
The error bars are obtained by fitting $\sigma_{12}$ separately from 
each of the jackknife samples and computing the associated jackknife 
error (analogous to the way we obtain errors on the power-law fit for 
$w_p$). This provides a realistic estimate of the errors, which are 
dominated by variations in the number of rare, high-dispersion structures 
in the sample (see discussions by 
\citealt{mo93,zurek94,marzke95,somerville97}). 
Figure~\ref{fig:xi_pi} compares the function $\xi(r_p,\pi)$ predicted
by the best fit model to the measured values for several different
choices of $r_p$.  

As a test of the sensitivity of our results to the assumed form of 
${\overline v_{12}}$, we have repeated the analysis where  
${\overline v_{12}}$ is assumed to be the same as that for dark 
matter particles in the $\Lambda$CDM model (calculated using the 
formulae in \citealt{mo97}). This assumption would be valid if the mean 
streaming velocity of galaxy pairs at a given separation is the same as 
that of mass particles at the same separation. We find that for
$r_p\la 3 \hmpc$, the PVD is quite similar to that obtained assuming
the similarity model (eq.~\ref{eq:v12}), while  at larger separations 
it changes significantly. Without knowing  how galaxies are biased 
relative to the mass, it is unclear which infall model is more realistic. 
The test we describe here, however, indicates that estimates of the PVD 
at $r_p \la 3 \hmpc$ are robust to uncertainties in the infall model.

The measured PVD is roughly constant in this range, 
with $\sigma_{12}(r) \simeq 550-675\kms$.  If we adopt an 
EdS cosmology, $\sigma_{12}$ decreases by 
$\approx 50-100\kms$. 
The last column in Table~\ref{table:comp} presents the values
of $\sigma_{12}$ at $r_p=1\hmpc$ obtained by our analysis and
some other redshift surveys.
Our estimate is close to the values found by \citet{jing98} for the LCRS, 
$570\kms$, and by \citet{marzke95} for CfA2+SSRS2, $540\kms$,
but it is substantially higher than the values found in the early 1980s 
from much smaller redshift surveys ($250\kms$ by \citealt{bean83};
$340\kms$ by \citealt{davis83}). The SDSS result thus confirms
that the galaxy velocity field, while colder than predicted by 
unbiased $\Omega_m=1$ models (\eg, \citealt{davis85}), is not so cold
that it demands an extremely low value of $\Omega_m$ or a highly
biased galaxy distribution.  While $\sigma_{12}(r)$ has been the most
widely used characterization of small-scale velocity dispersions,
other statistics have been proposed that are less sensitive to
rare objects that contribute many pairs \citep{davis97,landy98} or that 
quantify the dispersion as a function of local density \citep{strauss98}.
Future measurements that examine the dispersion as a function of both
environment and type may prove a valuable diagnostic for 
the relation between galaxies and dark matter halos \citep{sheth01b}.

\section{Dependence on Galaxy Properties}
\label{sec:dependencies}

The SDSS is ideal for investigating the dependence of clustering 
on galaxy properties because a wealth of photometric data is available 
for each galaxy in the spectroscopic sample.  
Here we examine the dependence of the real-space correlation
function and redshift-space anisotropy on galaxy color, then calculate
the real-space correlation function for subsamples defined by luminosity,
surface brightness, and light-profile concentration.
The spirit of our investigation is similar to that of Guzzo et al.'s
(\citeyear{guzzo97}) study of galaxy clustering as a function of
morphological type and luminosity, but the higher quality of our imaging data
allows us to consider a broader set of photometric parameters, 
and the larger size of our redshift sample allows us to measure
differences in clustering with higher precision.

\subsection{Color}
\label{subsec:color}

We divide our full sample into two subsamples based on the rest-frame
$\us-\rs$ colors of the galaxies.  \citet{strateva01} find that the $\us-\rs$
color distribution of galaxies is bimodal, and thus galaxies can be 
naturally divided into ``blue'' and ``red'' classes. They also show 
(using independent morphological classification schemes) that the blue 
class contains mainly late (spiral) morphological types while the red 
class consists mainly of bulge-dominated galaxies, as one would expect.
After $K$-corrections are accounted for, we find that the color distribution 
is still bimodal but that the division at $\us-\rs=2.2$ in observed bands 
is closer to $\us-\rs=1.8$ in rest-frame bands. 
We therefore divide the sample into galaxies bluer and redder than a
rest-frame color of $\us-\rs=1.8$, resulting in a red subsample that
includes $\sim 20,000$ galaxies and a blue subsample of $\sim 10,000$
galaxies. In the full absolute magnitude range considered here 
($-19<M_{\rs}<-22$), the two subsamples have similar space densities,
but the red galaxies are systematically more luminous \citep{blanton01a}
and therefore sample a larger volume.  We list some relevant properties
of the full sample and the color subsamples in the first lines of
Table~\ref{table:samples}.  Space densities $\bar n$ are the inferred
mean density for the indicated class of galaxies over the
full absolute magnitude range.  
We also repeated our clustering analysis defining the blue and red samples
based on the rest-frame $\gs-\rs$ color (making the division at
$\gs-\rs=0.6$), and found very similar results.

\begin{deluxetable}{lcccccc}
\tablewidth{0pt}
\tablecolumns{7}
\tablecaption{\label{table:samples} Flux-limited Correlation Function Samples}
\tablehead{ Description & Additional
limits & $N_{\mathrm{gal}}$ & ${\bar n}$ & $r_0$ & $\gamma$ & 
$r_{r_0\gamma}$ }
\tablecomments{All samples use $14.5 < r^\ast <17.6$, $5,700 \kms < cz
< 39,000 \kms$, and $-22 < M_{r^\ast} < -19$.
${\bar n}$ is measured in units of $10^{-2}$ $h^{3}$ Mpc$^{-3}$.
$r_0$ is in units of $\hmpc$.
$r_0$ and $\gamma$ are obtained from a fit for $w_p(r_p)$.
$r_{r_0\gamma} \equiv \sigma_{r_0\gamma}/ 
\sqrt{\sigma_{r_0}\sigma_{\gamma}}$ is the correlation coefficient 
between $r_0$ and $\gamma$.
} 
\startdata
Full & --- & 29,300 & 1.85 & 6.14 $\pm$ 0.18 & 1.75 $\pm$ 0.03 & -0.51 \cr
Red & $u-r>1.8$ & 19,603 & 1.05 & 6.78 $\pm$ 0.23 & 1.86 $\pm$ 0.03 & -0.15 \cr
Blue & $u-r<1.8$ & 9,532 & 0.87 & 4.02 $\pm$ 0.25 & 1.41 $\pm$ 0.04 & -0.24 \cr
High SB & $\mu_{1/2,r^\ast} < 20.5$ & 17,859 & 0.94 & 6.48 $\pm$ 0.21 &
1.84 $\pm$ 0.03 & -0.14 \cr
Low SB & $\mu_{1/2,r^\ast} > 20.5$ & 11,439 & 0.98 & 5.55 $\pm$ 0.21 &
1.55 $\pm$ 0.04 & -0.47 \cr
High Concen.\ & $c=r_{90}/r_{50}>2.7$ & 11,883 & 0.55 & 6.74 $\pm$ 
0.24 & 1.88 $\pm$ 0.02 & -0.29 \cr
Low Concen.\ & $c=r_{90}/r_{50}<2.7$ & 17,417 & 1.41 & 5.64 $\pm$
0.22 & 1.63 $\pm$ 0.03 & -0.01 \cr
\enddata
\end{deluxetable}

\begin{figure}[tbp]
{\includegraphics{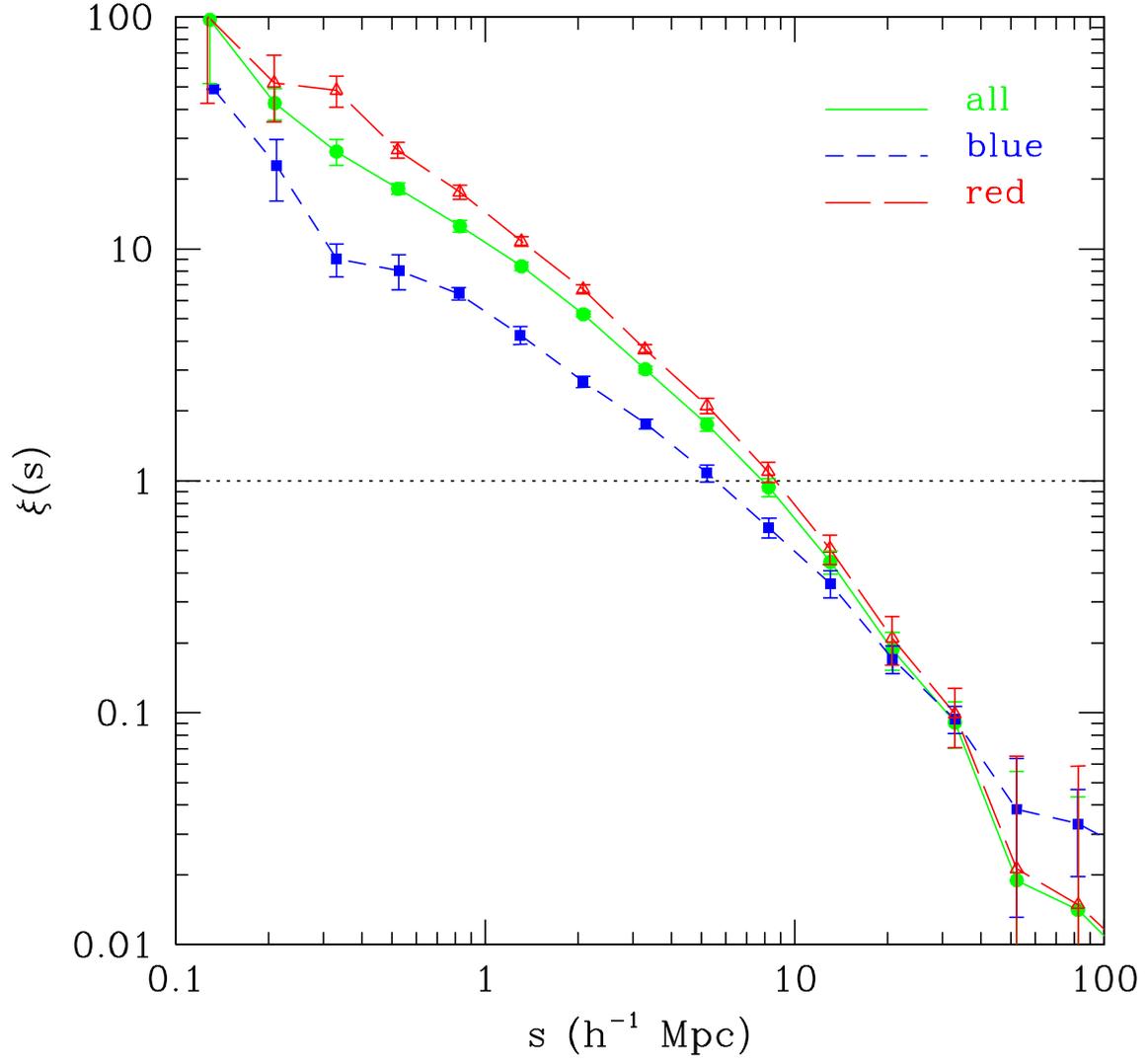}}
\vspace{16 cm}
\caption{The redshift-space correlation function $\xi(s)$ for the blue 
sample (solid squares, short-dashed line), the red sample (open triangles, 
long-dashed line) and the full sample (solid circles and line). Color cut 
is based on $\us-\rs$ color.}
\label{fig:xsis_br}
\end{figure}

\begin{figure}[tbp]
{\includegraphics{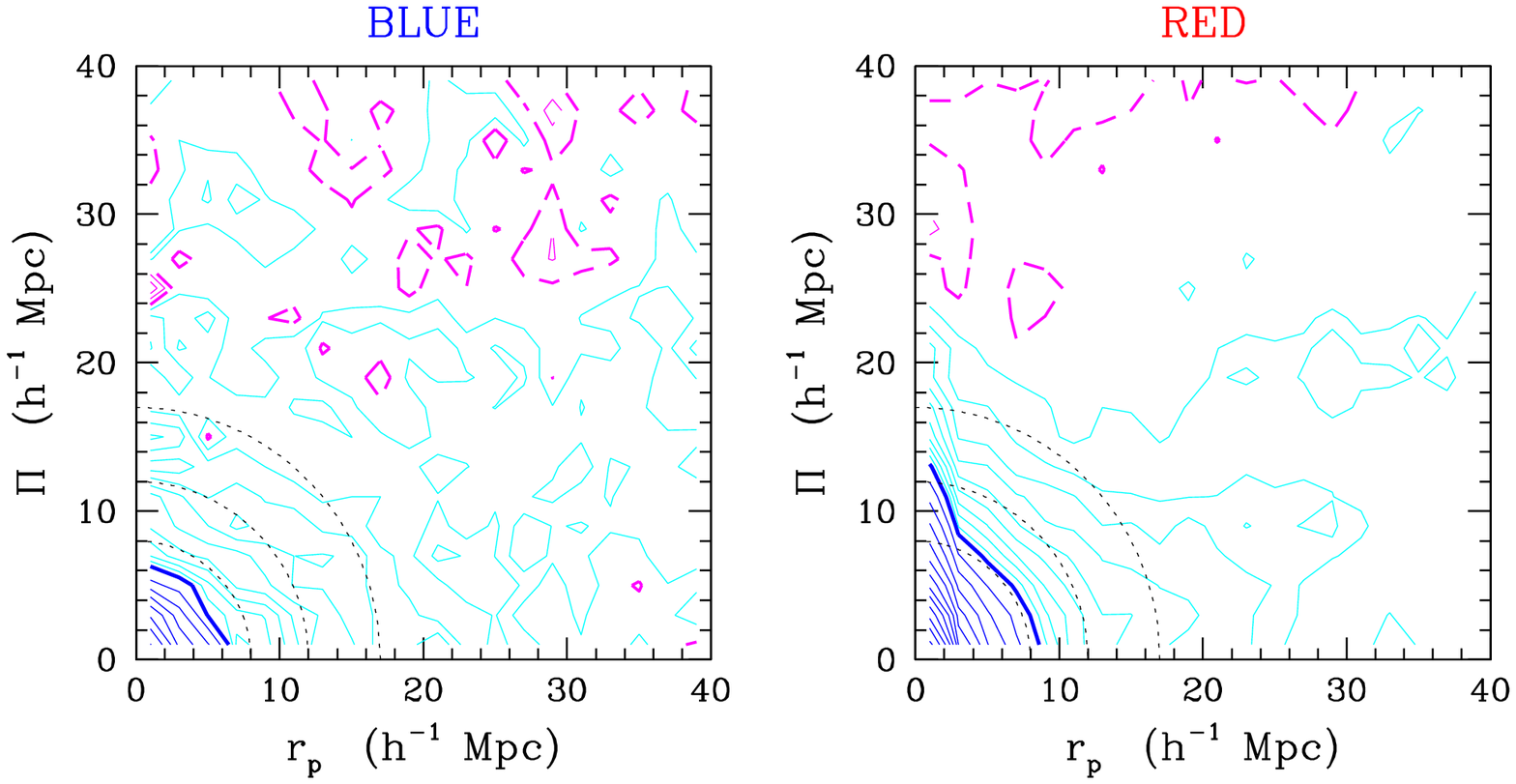}}
\vspace{13 cm}
\caption{$\xi(r_p,\pi)$ for the blue sample (left panel) and red sample
(right panel). Contours are as in Fig.~\ref{fig:xsirpi}.}
\label{fig:xsirpi_br}
\end{figure}

Figure~\ref{fig:xsis_br} compares the redshift-space correlation functions
of the red and blue subsamples to that of the full galaxy sample.
The red galaxies have a substantially higher amplitude and steeper
$\xi(s)$ than the blue galaxies, with a correlation length 
$s_0\approx 9\hmpc$ compared to $s_0 \approx 5.5\hmpc$ for the blue
galaxies.  This difference is expected from
the well known morphology-density relation \citep{dressler80}, 
since redder, early-type
galaxies preferentially inhabit high density regions.
The difference in anisotropy of $\xi(r_p,\pi)$ is equally 
striking (Fig.~\ref{fig:xsirpi_br}), with red galaxies exhibiting
much stronger finger-of-God distortions on small scales.
The compression of contours along the $\pi$ axis at large scales
is also much more obvious for the red galaxies, though this may
be just a consequence of the smaller number and weaker clustering
of the blue galaxies, which makes $\xi(r_p,\pi)$ much noisier.

\begin{figure}[tbp]
{\includegraphics{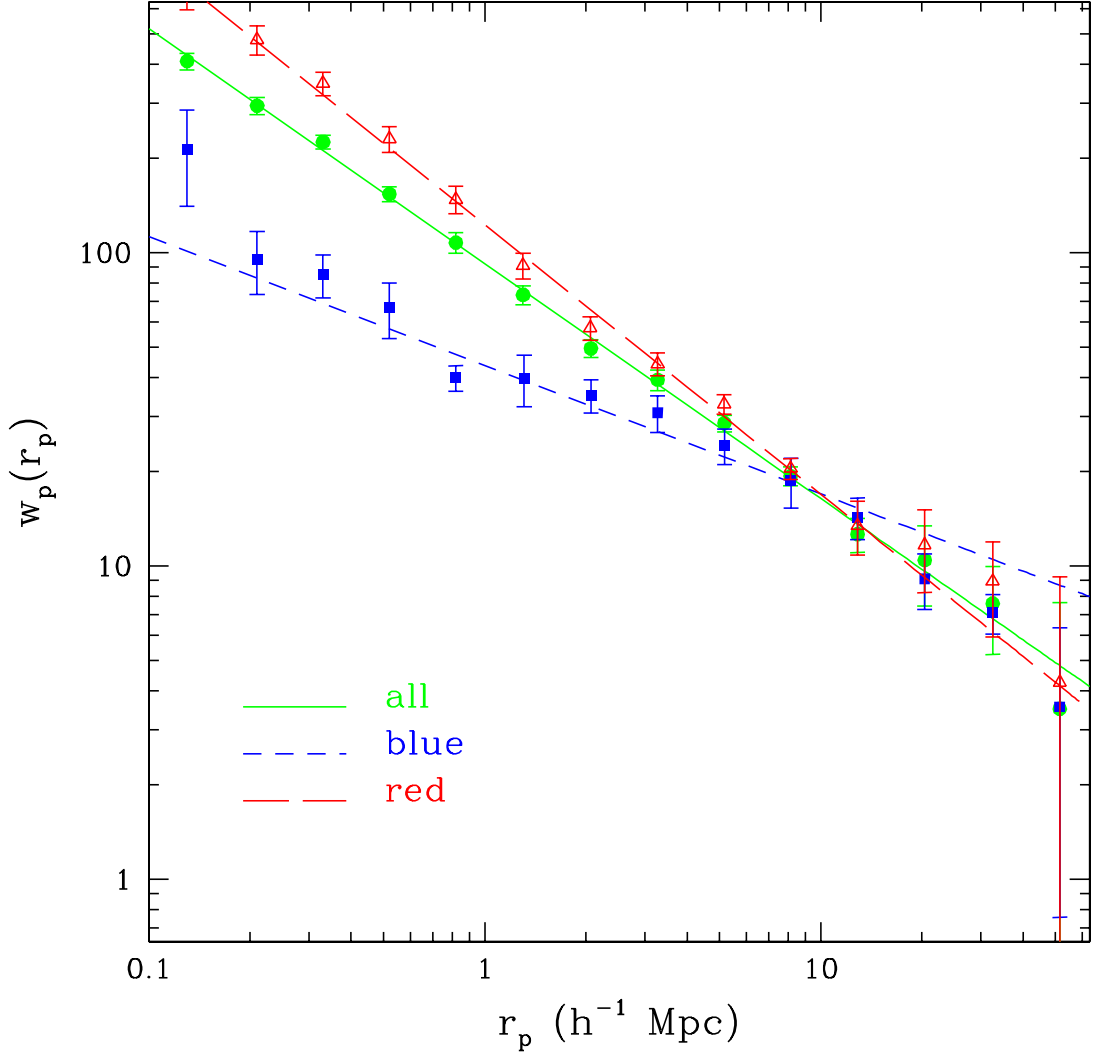}}
\vspace{16 cm}
\caption{Projected correlation functions $w_p(r_p)$ for the blue (squares)
red (triangles), and full (circles) samples.
The straight lines are the best-fit  power-laws for $w_p$, obtained for
$0.1\hmpc<r_p<16\hmpc$. The short-dashed, long-dashed and solid lines
correspond to the blue, red, and full samples, respectively. }

\label{fig:wp_br}
\end{figure}

Because the peculiar velocity distortions are very different for
the two subsamples, it is important to remove them in order to assess
their relative spatial clustering.
Figure~\ref{fig:wp_br} compares the projected correlation functions
$w_p(r_p)$, with the red galaxies again exhibiting a steeper and
higher amplitude correlation function.  Power-law model fits in the
range $0.1\hmpc < r_p < 16\hmpc$ yield 
$r_0=6.78\pm 0.23\hmpc$, $\gamma=1.86\pm 0.03$ for the red galaxies and 
$r_0=4.02\pm 0.25\hmpc$, $\gamma=1.41\pm 0.04$ for the blue galaxies.
The blue galaxies show hints of a departure from power-law behavior
at the smallest separations, a possible signature of their tendency
to cluster in lower mass halos with smaller virial radii
(see, \eg, \citealt{seljak00}), but the statistical significance of
this departure is not high.  At large scales, the two correlation 
functions approach each other, with $w_p(r_p)$ for the red galaxies 
having a slightly higher amplitude but similar shape.  The behavior
in Figure~\ref{fig:wp_br} is qualitatively consistent with expectations
based on the morphology-density relation (Narayanan, Berlind, \& Weinberg
2000, figure 2), 
though the data at large scales are too noisy to test whether the relative 
bias becomes constant in the linear regime, as ``local'' bias models
predict \citep{coles93,fry93,mann98,scherrer98,narayanan00}.

\begin{figure}[tbp]
{\includegraphics{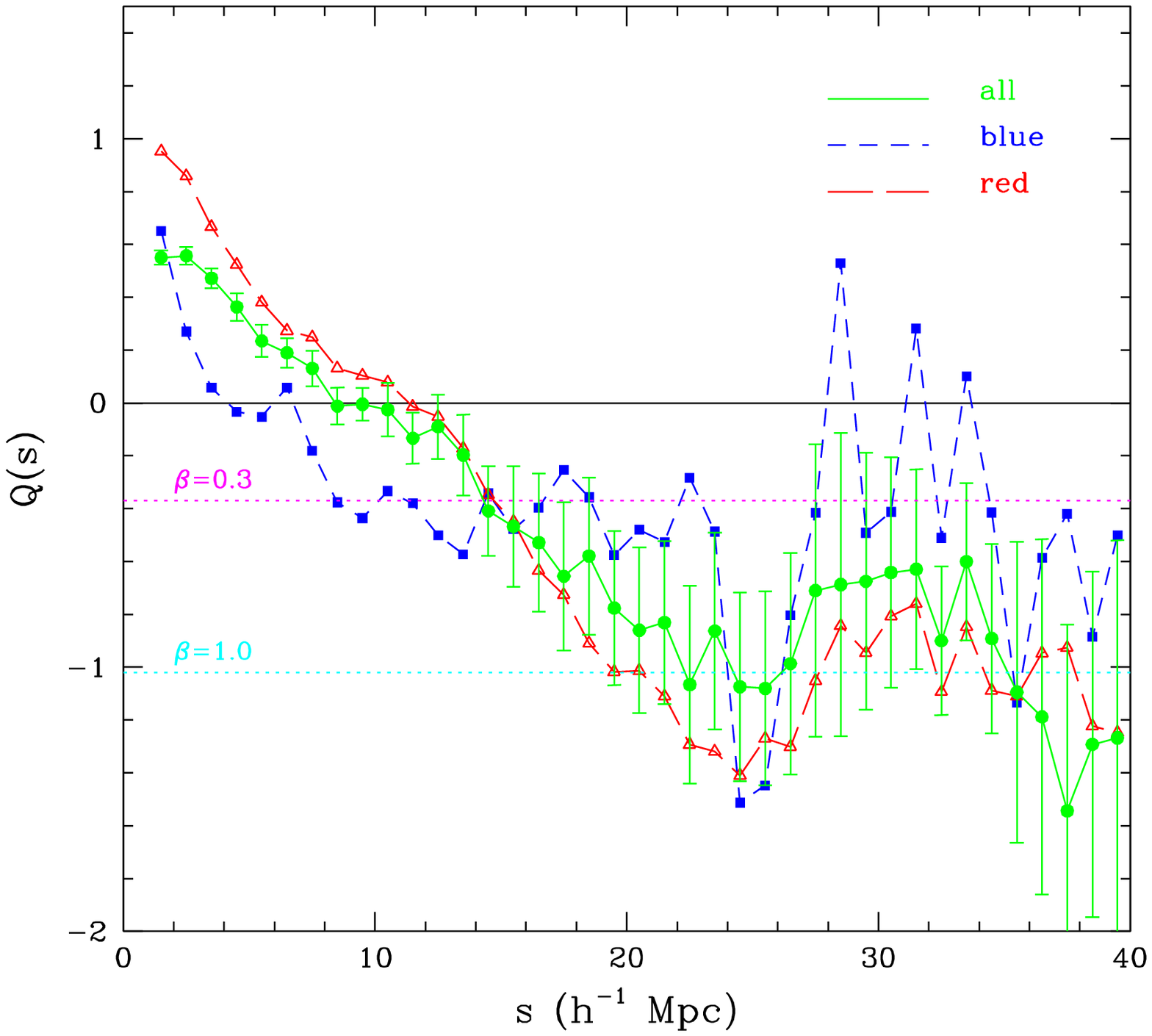}}
\vspace{16 cm}
\caption{$Q(s)$ for the blue (short-dashed), red (long-dashed) and 
full (solid) samples. For clarity, error bars are drawn only for
the full sample. }
\label{fig:qlin_br}
\end{figure}

Figure~\ref{fig:qlin_br}, the quadrupole ratio $Q(s)$, confirms
the much stronger finger-of-God distortion of the red galaxies
evident in Figure~\ref{fig:xsirpi_br}, with a large positive $Q$
at small scales.  At large scales, the red
galaxies have a more negative $Q(s)$ than the blue galaxies, which
is contrary to expectation given their higher relative bias, but
the difference is marginal at best; for $s>15\hmpc$, both subsamples
generally have $Q(s)$ within the $1\sigma$ error bar of the full
sample $Q(s)$.  With future, larger samples, comparison of real-space
clustering amplitudes and redshift-space distortions on large scales
will allow interesting new tests of bias models.

\begin{figure}[tbp]
{\includegraphics{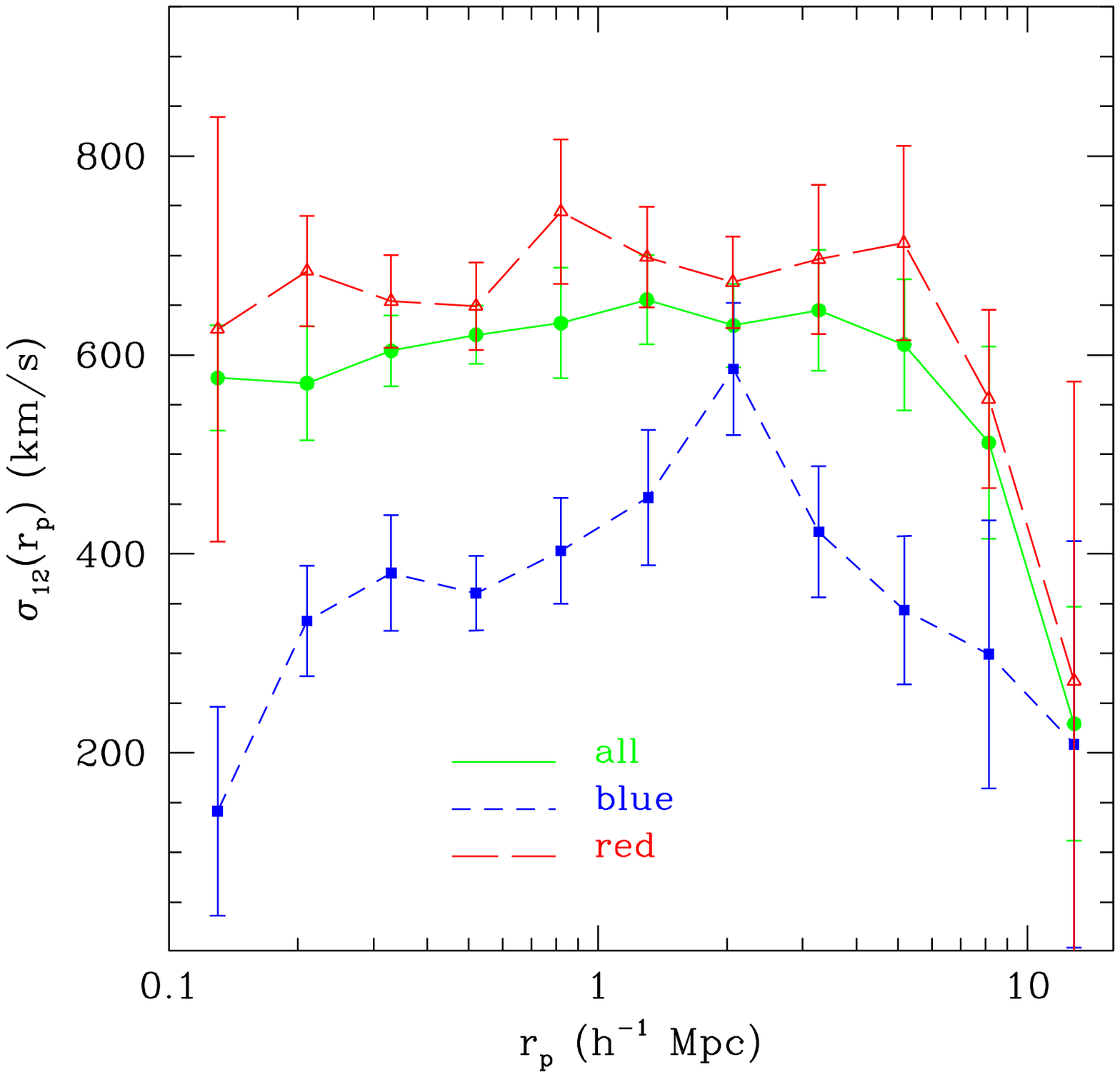}}
\vspace{16 cm}
\caption{The pairwise velocity dispersion $\sigma_{12}(r_p)$ for the blue
(short-dashed), red (long-dashed), and full (solid) samples. All error bars 
are $1\sigma$, derived from jackknife subsamples of the indicated galaxy class.
}
\label{fig:pvd_br}
\end{figure}

Figure~\ref{fig:pvd_br} shows the pairwise velocity dispersions of 
the two subsamples, demonstrating very clearly the preference of
red galaxies for denser, hotter environments.  For $r_p \sim 0.2-8\hmpc$,
the PVD of the red sample is $\sigma_{12}\approx 650-750\kms$, while
the blue galaxy PVD is only $\sigma_{12}\approx 300-450\kms$.
This latter range is in fact similar to that obtained by \citet{fisher94}
for IRAS galaxies. 
The amplitude and scale-dependence we obtain for the blue sample agree 
as well with a similar calculation by \citet{jing01} using the PSCz survey.
Our two subsamples have similar $\sigma_{12}$ at 
$r_p=15\hmpc$; partial convergence of $\sigma_{12}$ at large scales is 
expected in theoretical models \citep{sheth01a}, though the assumptions 
used to infer $\sigma_{12}$ from $\xi(r_p,\pi)$ may also be breaking down 
at this point (see \S\ref{subsec:pvd} and \citealt{fisher95a}).
Calculations where ${\overline v_{12}}$ is assumed to be the same as that 
for dark matter particles in the $\Lambda$CDM model show again that the PVD 
at $r_p\la 3 \hmpc$ are quite robust against the change of infall model. 

\subsection{Luminosity}
\label{subsec:luminosity}

We study the dependence of clustering on luminosity using three
volume-limited subsamples, each with different absolute magnitude
and redshift limits, as summarized in Table~\ref{table:vlsamples}.
The absolute magnitude ranges of the three subsamples are centered
approximately on $M_*+1.5$, $M_*$, and $M_*-1.5$, where $M_*=-20.8$
is the characteristic luminosity in a \citet{schechter76} function
fit to the SDSS luminosity function 
(\citealt{blanton01a}; the other parameters in the fit are 
$\alpha=-1.2$ and $\phi_*=1.46 \times 10^{-2} h^{3}$ Mpc$^{-3}$). 
The space density of the lowest luminosity subsample is 27 times
that of the highest luminosity subsample. The redshift ranges are chosen 
to ensure that the selection function $\phi(z)=1$ for each subsample
(i.e., they are volume-limited),  accounting for $K-$corrections and the
bright and faint apparent magnitude limits of the full sample.
Because the width of the absolute magnitude bins is half the range of 
apparent magnitudes in the full sample, the three redshift ranges
are actually disjoint, so our comparison of clustering properties relies 
on each subsample volume being large enough to fairly represent
the cosmic mean.

\begin{deluxetable}{ccccccc}
\tablewidth{0pt}
\tablecolumns{7}
\tablecaption{\label{table:vlsamples} Volume-limited Correlation
Function Samples}
\tablehead{ Absolute Mag.\ Limits & Redshift Limits
& $N_{\mathrm{gal}}$ & ${\bar n}$ & $r_0$ & $\gamma$ & $r_{r_0\gamma}$ }
\tablecomments{All samples use $14.5 < r^\ast <17.6$.
${\bar n}$ is measured in units of $10^{-2}$ $h^{3}$ Mpc$^{-3}$.
$r_0$ is in units of $\hmpc$.
$r_0$ and $\gamma$ are obtained from a fit for $w_p(r_p)$.
$r_{r_0\gamma}$ is the normalized correlation coefficient between 
$r_0$ and $\gamma$.
}
\startdata
$-23.0 < M_{r^\ast} < -21.5$ & $0.100 < z < 0.174$ & 3,674 & 0.06 & 
7.42 $\pm$ 0.33 & 1.76 $\pm$ 0.04 & -0.85 \cr
$-21.5 < M_{r^\ast} < -20.0$ & $0.052 < z < 0.097$ & 9,067 & 0.73 & 
6.28 $\pm$ 0.77 & 1.80 $\pm$ 0.09 & -0.77 \cr
$-20.0 < M_{r^\ast} < -18.5$ & $0.027 < z < 0.051$ & 3,130 & 1.64 & 
4.72 $\pm$ 0.44 & 1.86 $\pm$ 0.06 & -0.83 \cr
\enddata
\end{deluxetable}

\begin{figure}[tbp]
{\includegraphics{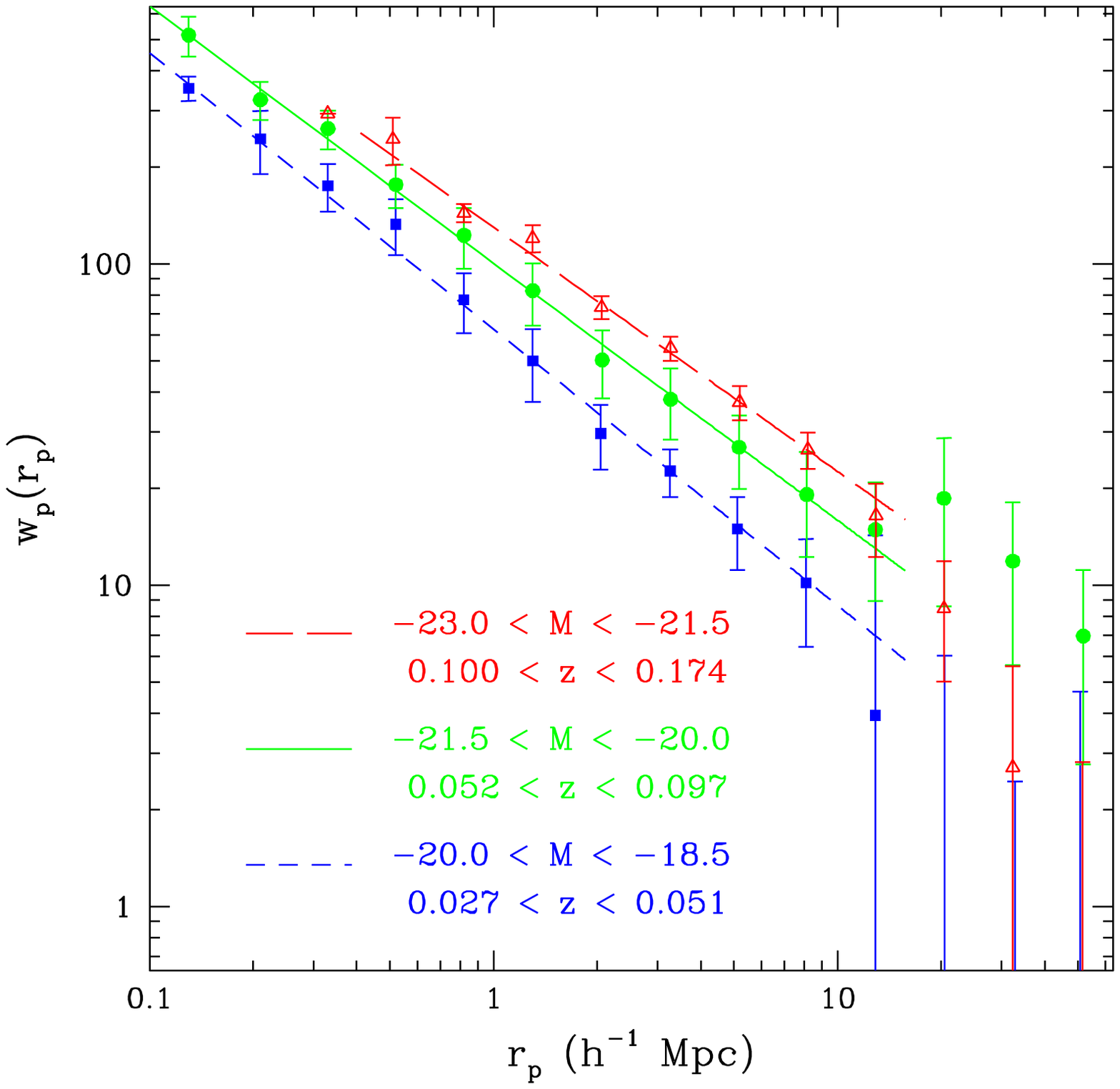}}
\vspace{16 cm}
\caption{Projected correlation function $w_p(r_p)$ for three volume-limited 
samples, with absolute magnitude and redshift ranges as indicated.
Squares, circles, and triangles show results for faint (sub-$M_*$), 
intermediate ($M_*$), and luminous (super-$M_*$) galaxies, respectively.
Short-dashed, solid, and long-dashed lines show the best fit power-law
models in the range they were fitted; parameters of the corresponding 
real-space $\xi_r(r)$ appear in Table~\ref{table:vlsamples}.  
}
\label{fig:wp_vl}
\end{figure}

Figure~\ref{fig:wp_vl} shows the projected correlation functions
$w_p(r_p)$ for the three absolute magnitude subsamples.
Table~\ref{table:vlsamples} lists the parameters $r_0$ and $\gamma$
of power-law $\xi_r(r)$ models determined by fitting $w_p(r_p)$ in
the range $0.4\hmpc < r_p < 16\hmpc$ for the highest luminosity
subsample and $0.1\hmpc < r_p < 16\hmpc$ for the other two subsamples.  
The correlation length and slope of the middle sample is similar to that
of the full sample analyzed in \S\ref{sec:results}, which is not
surprising since most of the galaxies in a flux-limited sample have
absolute magnitudes in the neighborhood of $M_*$.  The low luminosity
subsample has a clustering amplitude that is lower by $\sim 40\%$,
and the high luminosity subsample has a clustering amplitude higher
by $\sim 35\%$.

The general trend of Figure~\ref{fig:wp_vl}, stronger clustering for
more luminous galaxies, is similar to that found in a number
of earlier studies
\citep{davis88,hamilton88,white88,park94,loveday95,guzzo97,benoist98,willmer98,norberg01}.
However, while some of these studies (including Norberg et al.'s analysis 
of a large sample from the 2dFGRS) found that luminosity dependence
became strong only for galaxies brighter than $M_*$, we find a steady 
trend from $M_*+1.5$ to $M_*-1.5$.  Most of the earlier studies were
based on $B$-band luminosities, while we have used $\rs$-band luminosities,
and since clustering is color-dependent, this difference may partly or 
fully explain the difference in trend (see also \citealt{shepherd01}). 
As the SDSS sample grows, we will be able to examine luminosity 
dependence of clustering in greater detail over a wider dynamic range.

Perhaps the most remarkable aspect of Figure~\ref{fig:wp_vl} is the
nearly identical {\it shape} of the three correlation functions,
\ie ``scale-independent luminosity bias''; 
at the $1\sigma$ level each is consistent with a power-law $\xi_r(r)$
of slope $\gamma=1.8$ (see Table~\ref{table:vlsamples}).
The 2dFGRS analysis \citep{norberg01} also recovers nearly identical 
power-law slopes in the different luminosity ranges. 
A ``halo occupation'' analysis of galaxy clustering
(see, \eg, \citealt{benson99,ma00,peacock00,seljak00,berlind01,scoccimarro01})
implies that the correlation function at sub-Mpc scales is
dominated by pairs of galaxies that reside in the same virialized dark halo,
while the correlation function at scales $\ga 2\hmpc$ comes from
pairs in separate halos.  Maintaining the constant slope seen in 
Figure~\ref{fig:wp_vl} requires maintaining the relative 
strength of these two contributions in galaxy populations that differ
by a factor of 27 in space density and a factor of $2.3$ in correlation
amplitude, a delicate balancing act.  This empirical result should
prove a demanding constraint for theoretical models of galaxy formation.

\subsection{Dependence on Surface Brightness and Morphology}
\label{subsec:other}

The SDSS photometric pipeline (\citealt{lupton01a}, \citeyear{lupton01})
measures many
other properties that can be used to define galaxy classes.
Here we consider two of these properties, surface brightness and
light-profile concentration. 
Contrary to the case of the $\us-\rs$ color cut, where the bimodal 
distribution provides a natural place to divide the sample, for 
both these properties there isn't an obvious place to cut, so our 
division is somewhat arbitrary in the middle of the distribution.  
Table~\ref{table:samples} summarizes the thresholds that we use to 
define surface-brightness and concentration subsamples, along with 
sample sizes, mean space densities, and correlation function parameters.

\begin{figure}[tbp]
{\includegraphics{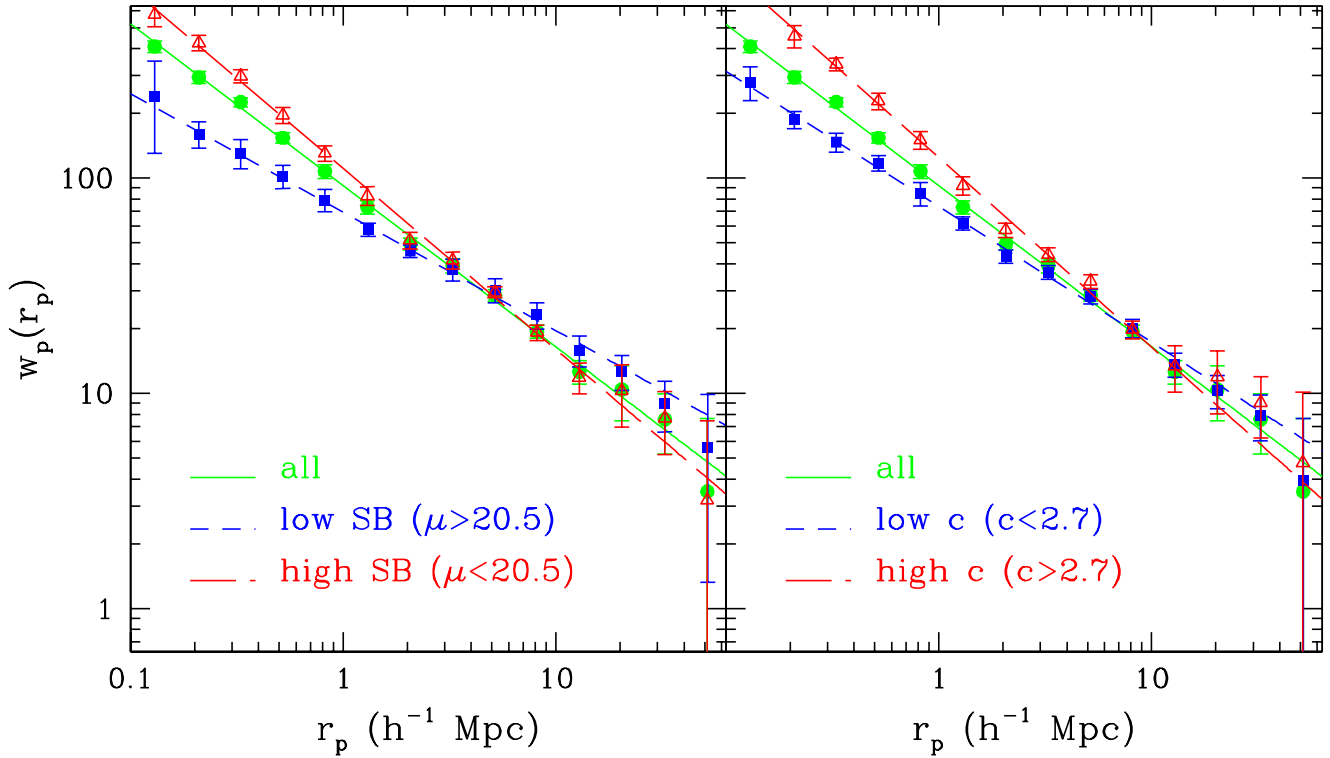}}
\vspace{13 cm}
\caption{Projected correlation function $w_p(r_p)$ for samples cut 
according to surface brightness $\mu$ (left panel) and 
concentration parameter $c \equiv r_{90}/r_{50}$ (right panel).
The straight lines correspond to the power-law fits of $w_p(r_p)$,
for $r_p < 16\hmpc$.
} 
\label{fig:wp_misc}
\end{figure}

The surface-brightness subsamples are divided at the threshold
$\mu_{1/2} = 20.5$ mag arcsec$^{-2}$,
where $\mu_{1/2} = m + 2.5 \log_{10} (2\pi r_{50}^2)$
is the mean $\rs$ surface brightness within the
Petrosian half-light radius $r_{50}$, $K$-corrected and corrected for
cosmological surface-brightness dimming. The low surface-brightness
sample contains around 11,400 objects and the high surface-brightness
sample contains around 17,900 objects. The left panel of 
Figure~\ref{fig:wp_misc} shows the projected correlation functions $w_p(r_p)$
of the two subsamples and of the full sample.
The high surface-brightness galaxies have a steeper $w_p(r_p)$ and a
higher clustering amplitude at $r_p \la 3\hmpc$.  
Fits of a power-law $\xi_r(r)$ to points with $r_p<16\hmpc$ yield
$r_0=5.55 \pm 0.21 \hmpc$, $\gamma=1.55 \pm 0.04$ 
for the low surface-brightness sample and
$r_0=6.48 \pm 0.21 \hmpc$, $\gamma=1.84 \pm 0.03$ for the high
surface-brightness sample. This trend of clustering strength with surface 
brightness is consistent with some earlier results based on smaller 
samples (\citealt{bothun93}, \citealt{mo94}).
The two correlation functions actually cross at large scales, 
contrary to the expectation from simple bias models 
(see \citealt{narayanan00}), but the $w_p(r_p)$ amplitudes on 
these scales are consistent with each other at the $1\sigma$ level.
We also note that since $w_p(r_p)$ is an integral in the $\pi$ direction 
out to $40 \hmpc$, its value at $r_p$ in fact probes clustering out to 
considerably larger scales. 

The profile concentration subsamples are defined
using the concentration parameter $c \equiv r_{90}/r_{50}$,
which serves as a proxy for the traditional division of
galaxies into early and late morphological types.
For example, a pure de~Vaucouleurs profile has $c \approx 3.3$ 
(given our definition of Petrosian magnitudes; see \citealt{blanton01a}), 
while a pure exponential profile has $c \approx 2.3$. We divide our full 
galaxy sample roughly in between these two values 
at $c=2.7$, yielding about 11,900 galaxies with high 
concentration and 17,400 galaxies with low concentration. The right 
panel of Figure~\ref{fig:wp_misc} shows the projected correlation 
functions of these subsamples.  As expected from earlier studies of 
morphology-dependent clustering (e.g.\ \citealt{guzzo97} and references 
therein), 
high-$c$ (early type) galaxies have a steeper, higher amplitude correlation 
function.  Fits of a power-law $\xi_r(r)$ yield 
$r_0=6.74 \pm 0.24 \hmpc$, $\gamma=1.88 \pm 0.02$ for the high
concentration subsample and 
$r_0=5.64 \pm 0.22 \hmpc$, $\gamma=1.63 \pm 0.03$ for the low concentration 
subsample.

Qualitatively, our results for galaxy subsamples defined by surface
brightness or profile concentration parallel our results for color
subsamples described in \S\ref{subsec:color}.  We have focused on
$w_p(r_p)$ and $\xi_r(r)$, but the same characterization extends to
redshift-space anisotropy: like red galaxies, high surface-brightness
and high concentration galaxies show strong finger-of-God distortions,
which in turn imply high pairwise velocity dispersions.  Given the
well known correlations between galaxy morphology, color, and surface
brightness, these similarities are not surprising.  As the SDSS progresses,
it will be possible to extend this type of analysis to a much finer
level; for example, seeing if surface-brightness effects can be separated
from color effects, isolating extreme classes of low surface-brightness
or compact galaxies, comparing the clustering of galaxies with high
and low profile concentration at fixed half-light surface brightness,
or even comparing the clustering of barred and unbarred spirals or
``disky'' and ``boxy'' ellipticals.  Detailed clustering studies of
this sort should help disentangle the roles of early formation
history and late-time transformation in determining galaxy properties.

\section{Conclusions}
\label{sec:conclusion}

We have presented the first measurements of galaxy clustering from 
early SDSS spectroscopic data, based on a sample of $29,300$ galaxies. 
Since this sample covers a limited volume, spanning 
only $\sim 7$\% of the total projected survey area of the SDSS, 
our analysis has focused mainly on small-scale clustering. 
The sample used for this analysis has been chosen with care: in 
addition to a uniform flux limit imposed at bright and faint magnitudes, 
$14.5< r^* < 17.6$, the sample is limited in radial velocity, 
$5,700 \kms < cz < 39,000 \kms$, to avoid uncertainties 
introduced by evolution of the luminosity function, and in absolute 
magnitude, $-22 < M_{r^*} - 5 {\rm log}_{10}h < 19$, so that it  
is dominated by galaxies with $r^*$-band luminosities around $M_*$. 
While these cuts reduce the number of galaxies included in the sample 
by nearly a factor of two, they allow robust conclusions to be drawn 
from the measurements.  We have checked, for example, that our results 
are insensitive to details of the correlation function estimation, to 
uncertainties in the sample selection function, and to the effects of 
the 55-arcsecond minimum fiber separation. As discussed in Appendix A 
below, we have also tested our jackknife error estimation method 
using a large number of mock redshift catalogs drawn from N-body simulations 
of Cold Dark Matter models. These tests indicate that the jackknife errors 
used herein provide an accurate representation of the true statistical
uncertainties over  the scales of interest, at least in the context of 
these models.

For our full, flux-limited galaxy sample, we have measured the angle-averaged
redshift-space correlation function $\xi(s)$ and the two-dimensional
correlation function $\xi(r_p,\pi)$, projected the latter along the $\pi$
axis to infer the real-space correlation function
$\xi_r(r)$, measured angular moments to quantify the anisotropy induced
by peculiar motions, and modeled the small-scale anisotropy to infer the
galaxy pairwise velocity dispersion $\sigma_{12}(r)$.
Approximating the redshift-space correlation function by a power law,
$\xi(s)=(s/s_0)^{-\gamma}$, yields a correlation length
$s_0 \approx 8\hmpc$ and a slope $\gamma \approx 1.2$, but this
representation is not accurate over a large range of scales.
At small projected separations, contours of $\xi(r_p,\pi)$ show the 
characteristic ``fingers-of-God'' elongation along the line of sight caused 
by velocity dispersions in collapsed structures.  At large separations, they
show compression along the line of sight produced by coherent
flows into high density regions.  The projected correlation function
$w_p(r_p)$ can be well fit by a power-law real-space correlation function
$\xi_r(r)=(r/r_0)^{-\gamma}$, with $r_0=6.14 \pm 0.18 \hmpc$ and
$\gamma=1.75 \pm 0.03$, for projected separations from $0.1\hmpc$ 
to $30\hmpc$.  The ratio $Q(s)$ of the quadrupole
and monopole moments of $\xi(r_p,\pi)$ is positive for $s \la 10\hmpc$,
where ``finger-of-God'' distortions dominate,
and negative at larger scales, where coherent flow distortions dominate.
A future analysis using a larger sample that extends to large scales in 
all three dimensions  will enable us to extract an estimate of 
$\beta \equiv \Omega_m^{0.6}/b$ from the large-scale anisotropy.  From the 
elongation of $\xi(r_p,\pi)$ at small scales, we
estimate a pairwise velocity dispersion $\sigma_{12}(r) \approx 600\kms$
that is roughly constant in the range $0.1\hmpc < r < 10\hmpc$.

Our results for the full galaxy sample are in fairly good agreement
with those obtained from earlier optically selected galaxy redshift
surveys (see Table~\ref{table:comp}), in particular from clustering 
analyses of the LCRS \citep{tucker97,jing98}, which has similar selection,
geometry, and size. The fact that our first analysis of early
data from the SDSS reproduces these results and yields comparable or better
statistical precision demonstrates the encouraging prospects for future
galaxy clustering studies with the SDSS redshift survey.
If we restrict our analysis to the subset of galaxies included
in the SDSS Early Data Release \citep{stoughton01}, we obtain
very similar correlation function results, with larger statistical
uncertainties because the sample is about half the size.

Taking advantage of our large sample and the high quality of SDSS
imaging data, we have carried out a detailed examination
of the dependence of real-space correlations and redshift-space
distortions on galaxy photometric properties.
Red and blue galaxies display markedly different clustering statistics,
with the red galaxies exhibiting a higher amplitude and steeper real-space
correlation function and much stronger finger-of-God distortions
than the blue galaxies; at $r_p=1\hmpc$, the pairwise velocity dispersion 
is $\sim 750\kms$ for our red galaxy subsample and $\sim 350\kms$ for our 
blue galaxy subsample.
Subsamples of high/low surface brightness and high/low profile concentration
display qualitative behavior similar to that of the red/blue subsamples.
Perhaps our most striking result is a measurement of luminosity
bias of the real-space correlation function that is approximately
scale-independent at $r \la 10\hmpc$.
Using three volume-limited subsamples, we find a $\sim 40\%$ decrease
in clustering amplitude as we go from a median absolute magnitude of 
$M_*$ to $M_*+1.5$ and a similar increase when going from $M_*$ to $M_*-1.5$,
implying relative biasing parameters $b/b_* \equiv (\xi/\xi_*)^{0.5}$ of 
0.8 and 1.2, respectively, for the faintest and brightest samples. 
These three samples differ by a factor of 27
in galaxy number density, but in each case $\xi_r(r)$ is consistent
with a power law of slope $\gamma \approx 1.8$.

Studies of galaxy clustering and redshift-space distortions in the local
universe have two main scientific objectives:
(a) to test cosmological models and determine their parameters,
and (b) to infer the relation between the galaxy and dark matter distributions,
partly to sharpen cosmological tests, but mostly to constrain and guide
the emerging theory of galaxy formation.  Cosmological model tests
usually focus on large scales, where the effects of non-linear
gravitational evolution and biased galaxy formation are relatively simple.
These tests typically employ Fourier methods or statistical
techniques that can isolate large-scale information and produce 
approximately uncorrelated error estimates even in the presence of 
a complicated survey geometry \citep{vogeley96,tegmark97,tegmark98}.
Several of these methods have been applied to SDSS angular clustering
data \citep{tegmark01,dodelson01,szalay01}, and they will be applied
to the increasing sample of SDSS redshift data in the near future.
The best constraints on galaxy bias will probably come from small and
intermediate scales, where clustering statistics are most sensitive to
the relation between galaxies and dark matter and where precise measurements
can be obtained for many different classes of galaxies.
Our results on the color
and luminosity dependence of real-space clustering and on pairwise
velocities already provide a challenging target for theories
of galaxy formation. In the near future, these will be complemented
by measurements of higher-order clustering, which can break degeneracies
among bias models that match two-point correlations 
\citep{scoccimarro01,szapudi01}. Studies of galaxy-galaxy weak lensing 
in the SDSS offer an entirely new route to determining the
relation between galaxies and dark matter \citep{fischer00,mckay01}.
These measurements and other characterizations of galaxy clustering
will improve in precision and detail as the SDSS progresses, yielding
a wealth of new information with which to understand galaxy formation.

\acknowledgements{
IZ, MRB and JAF acknowledge support by the DOE and NASA grants NAG 5-7092 and
NAG 5-10842 at Fermilab and by NSF grant PHY-0079251 at UChicago. 

The Sloan Digital Sky Survey (SDSS) is a joint project of the University of 
Chicago, Fermilab, the Institute for Advanced Study, the Japan Participation
Group, the Johns Hopkins University, the Max-Planck-Institute for Astronomy
(MPIA), the Max-Planck-Institute for Astrophysics (MPA), New Mexico State 
University, Princeton University, the United States Naval Observatory, and 
the University of Washington. Apache Point Observatory, site of the SDSS 
telescopes, is operated by the Astrophysical Research Consortium (ARC). 

Funding for the project has been provided by the Alfred P.\ Sloan
Foundation, the SDSS member institutions, the National Aeronautics and Space
Administration, the National Science Foundation, the U.\ S.\ Department of
Energy, the Japanese Monbukagakusho, and the Max Planck Society. The SDSS Web 
site is http://www.sdss.org/.
}

\appendix
\label{app:jk}
\begin{center} {\bf A. Reliability of Error Estimates} \end{center}

A complete model of galaxy clustering predicts, in addition
to mean values, the distribution and covariance of statistical
measurement errors for any specified sample geometry and selection function.
These predicted statistical errors can be used to assess the consistency
of the model with the data.  However, such error estimates can be
cumbersome to compute, and they depend on the assumed clustering model itself.
For some purposes, therefore, it is desirable to have estimates of 
statistical errors and their covariances that depend only on the
data set itself.

One common approach to this task is to estimate errors from disjoint 
subsamples of the full data set, each occupying a separate sub-volume.
One calculates the statistic of interest --- \eg, $\xi(s_i)$ for
a number of separations $s_i$ --- in 
each sub-volume, and the estimated error of $\xi(s_i)$ is the
error on the mean determined from the $N$ sub-volumes.
The same approach can be used to estimate covariance of errors.
The disadvantage of this technique is that estimates of $\xi(s_i)$
from individual sub-volumes may become noisy or biased, especially
on scales comparable to the sub-volume size.
A related but more robust way to estimate errors from the sample itself is
the jackknife method described in 
Section~\ref{subsec:estimate} (see specifically eq.~\ref{eq:jk}).
In this approach, each jackknife subsample is obtained by {\it excluding}
one of the sub-volumes from the full sample, and one ``sums up'' the variances
of the jackknife subsamples rather than taking the error of the mean.
Since each jackknife subsample is similar in size to the full sample,
this method performs better on large scales, though the two approaches
should give equivalent results in the limit where each sub-volume
is representative of the whole data set (i.e., when fractional variances
are small).

Our error bars on plotted data points and on parameter estimates
($r_0$, $\gamma$, $\sigma_{12}$) are all computed using the jackknife
method.  Here we compare this approach to the model-based approach
using the mock SDSS catalogs of \citet{cole98}. \citet{cole98} ran a 
series of high resolution N-body simulations, using an Adaptive P$^3$M 
code \citep{couchman91}, for a suite of cosmological
models and biasing schemes. They created catalogs with the
survey geometry and anticipated selection function of the SDSS.
We use two of their catalogs: a COBE-normalized flat $\Lambda$CDM universe 
(with $\Omega_m=0.3$ and $h=0.65$) and a structure-normalized $\tau$CDM model 
(with $\Omega_m=1.0$, $h=0.5$ and $\Gamma=0.25$).  From each of these we 
extract $75$ galaxy samples, each of which resembles the observed stripes 
in our sample.  
For computational convenience, we used artificial samples that are smaller
than our current data set, with only $\sim 6,000$ galaxies per sample,
but we expect that our conclusions about the relative behavior of jackknife
and mock catalog error estimates would also hold for larger samples.

\begin{figure}[tbp]
{\includegraphics{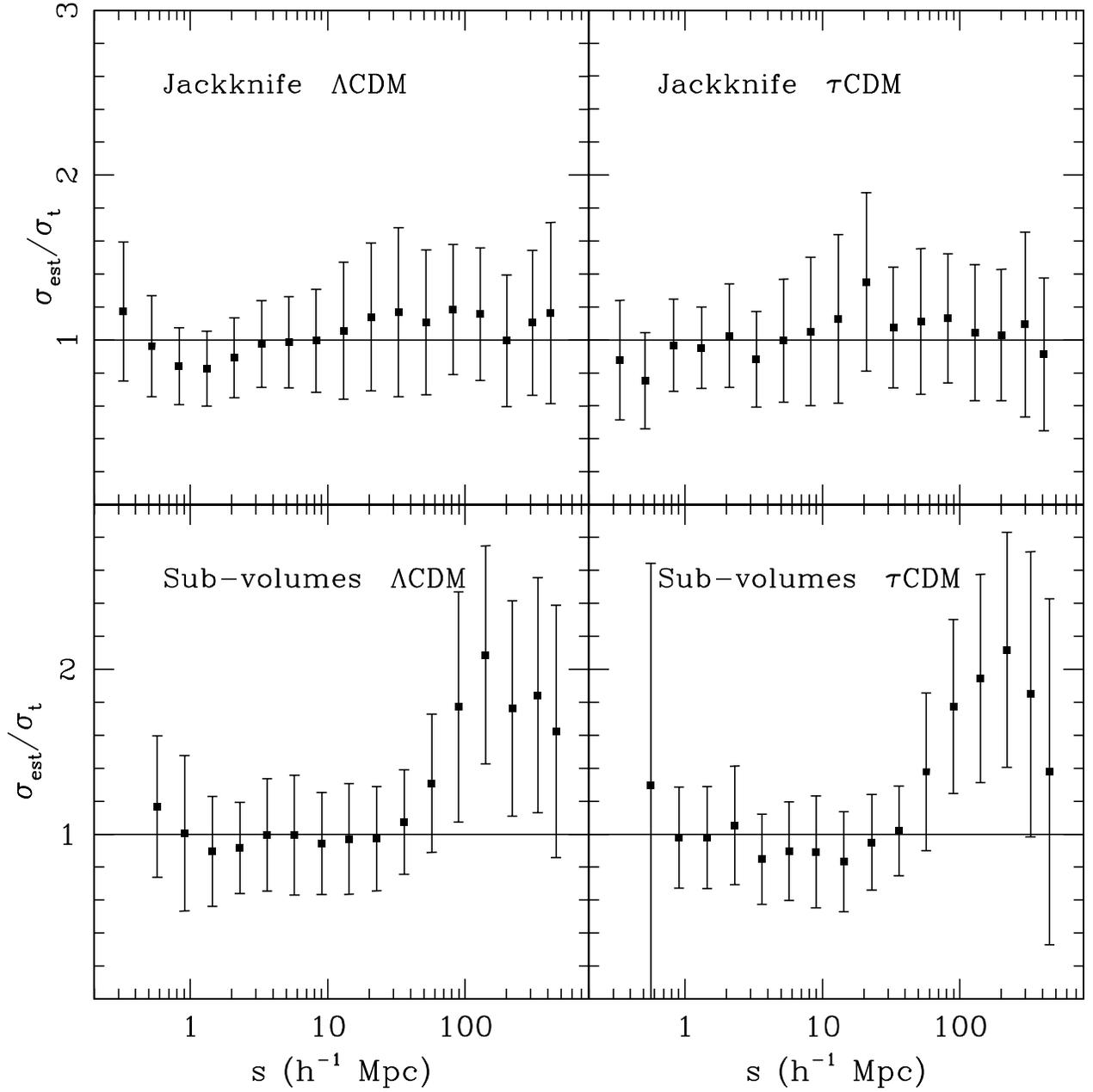}}
\vspace{17 cm}
\caption{Comparison of jackknife and sub-volume ``internal'' error
estimates to ``external'' estimates derived from variation among
artificial samples, using the N-body mock catalogs of \citet{cole98}.
Points and error bars show the mean and $1\sigma$
scatter of the error estimates derived from 75 artificial samples, 
in units of the true (external) error.
Top panels show results for the jackknife estimator and bottom panels
for the sub-volume estimator.  Left and right panels show
$\Lambda$CDM and $\tau$CDM models, respectively.
}
\label{fig:errors}
\end{figure}

For each artificial galaxy sample we calculate the redshift-space 
$\xi(s)$ and the error estimates (including the full covariance matrix) 
using the jackknife method and the sub-volume method.
Figure~\ref{fig:errors} compares these estimated errors to the ``true''
errors of this model, defined as the scatter of the 75 $\xi(s)$ estimates
from different samples.  Points and error bars show the mean and $1\sigma$
scatter of the ``internal'' error estimates (jackknife or sub-volume)
in units of the true, ``external'' error at the same separation $s$.

The jackknife estimates recover the true errors reasonably well
(with $1\sigma$ scatter $\la 50\%$), and they are robust on all scales, 
without any gross systematics. The sub-volume estimates do comparably 
well on the intermediate scales, but on large scales they overestimate 
the errors. The sub-volume estimates are also more numerically
unstable on small scales, where a single sub-volume may contain
few galaxy pairs.

The jackknife method produces unbiased estimates of the true errors
for both cosmological models.  These models are both designed to match
the observed APM correlation function, so their predicted errors are
also comparable, but they do differ by $\sim 20\%$ on some scales,
and the jackknife estimates seem to track these variations.
We also performed this analysis for the off-diagonal elements
of the covariance matrices. The covariance estimates do fairly well at
intermediate scales but less well at small and large separations.
The scatter in the estimates progressively increases for elements farther
away from the diagonal. 
We are therefore less confident in the usefulness of the jackknife
method for estimating a full covariance matrix; estimates of 
off-diagonal terms may be noisy and inter-dependent, making inversion
of the covariance matrix unstable.  When estimating parameters like
$r_0$, $\gamma$, and $\sigma_{12}$, therefore, we fitted values
for each jackknife subsample using only diagonal terms, then estimated
the error on the parameter by summing the variance of the estimates
in the jackknife subsamples (see \S\ref{subsec:rspace} and~\ref{subsec:pvd}).
This approach seems more reliable than using the full jackknife
covariance matrix itself.

For a specified clustering model, the mock catalog approach is probably
the best way to assess the consistency of the model with the data,
provided that one has an efficient way to create large numbers of 
mock catalogs of the necessary size.  (\citealt{scocc01} present a novel
method that should improve the computational practicality of this 
approach for large-scale surveys.)
However, the tests presented here give us confidence that our
jackknife error estimates should be representative of the true statistical
error bars for models that have clustering similar to that observed.
The jackknife approach is especially convenient when one breaks up 
the full sample into subsets that have different clustering properties,
as we have done in \S\ref{sec:dependencies}, since it will automatically
account for the influence of these clustering differences on the
statistical errors.  \citet{scranton01} have compared jackknife and
mock catalog errors in the context of angular clustering measurements
and reached similar conclusions.

{}


\clearpage

\begin{thebibliography}{}

\bibitem[Abell(1958)]{abell58}
Abell, G.\ O. 1958, \apjs, 50, 241

\bibitem[Bahcall \& Soneira(1983)]{bahcall83}
Bahcall, N.\ A., \& Soneira, R. 1983, \apj, 270, 20 

\bibitem[Bardeen et al.(1986)]{bardeen86}
Bardeen, J., Bond, J.\ R., Kaiser, N., \& Szalay, A.\ S. 1986,
\apj, 304, 15

\bibitem[Bean et al.(1983)]{bean83}
Bean, A.\ J., Ellis, R.\ S., Shanks, T., Efstathiou, G., \& Peterson, B.\ A. 
1983, \mnras, 205, 605

\bibitem[Benoist et al.(1998)]{benoist98}
Benoist, C., Cappi, A., da Costa, L.\ N., Maurogordato, S., Bouchet,
F.\ R., \& Schaeffer, R. 1998, \apj, 514, 563

\bibitem[Benson et al.(1999)]{benson99}
Benson, A.\ J., Cole, S., Frenk, C.\ S., Baugh, C.\ M., \& Lacey, C.\ G. 2000,
\mnras, 311, 793

\bibitem[Berlind \& Weinberg(2001)]{berlind01}
Berlind, A.\ A., \& Weinberg, D.\ H. 2001, \apj, submitted, 
astro-ph/0109001

\bibitem[Blanton et al.(1999)]{blanton99}
Blanton, M.\ R., Cen, R., Ostriker, J.\ P., \& Strauss, M.\ A. 1999, \apj, 
522, 590

\bibitem[Blanton et al.(2001a)]{blanton01a} 
Blanton, M.\ R., et al. 2001a, \aj, 121, 2358

\bibitem[Blanton et al.(2001b)]{blanton01b} 
Blanton, M.\ R., Lupton, R.\ H., Maley, F.\ M., Zehavi, I., \& Loveday, J.
2001b, \aj, submitted, astro-ph/0105535

\bibitem[Bolzonella, Miralles \& Pell\'o(2000)]{bolzonella00}
Bolzonella, M., Miralles, J.-M., Pell\'o, R.~2000, A \& A, 363,476

\bibitem[Bothun et al.(1993)]{bothun93}
Bothun G.\ D., Schombert J.\ M., Impey C.\ D., Sprayberry D., 
McGaugh S.\ S. 1993, \aj, 106, 548

\bibitem[Cen \& Ostriker(1992)]{cen92}
Cen, R., \& Ostriker, J.\ P. 1992, \apj, 399, L113

\bibitem[Cole, Fisher \& Weinberg(1994)]{cole94}
Cole, S., Fisher, K., \& Weinberg, D.\ H. 1994, \mnras, 267, 785

\bibitem[Cole et al.(1998)]{cole98}
Cole, S., Hatton, S., Weinberg, D.\ H., \& Frenk C.\ S. 1998,
1998, 300, 945

\bibitem[Coleman, Wu, \& Weedman(1980)]{coleman80}
Coleman, G.\ D., Wu, C.-C., \& Weedman, D.\ W. 1980, \apjs, 43, 393

\bibitem[Coles(1993)]{coles93}
Coles, P., 1993, \mnras, 262, 1065

\bibitem[Col\'{\i}n et al.(1999)]{colin99}
Col\'{\i}n, P., Klypin, A.\ A., Kravtsov, A.\ V., \& Khokhlov, A.\ M. 1999,
\apj, 523, 32

\bibitem[Connolly et al.(2001)]{connolly01}
Connolly, A.\ J., et al. 2001, \apj, submitted,
astro-ph/0107417

\bibitem[Couchman(1991)]{couchman91}
Couchman, H.\ M.\ P. 1991, \apj, 368, 23

\bibitem[Csabai et al.(2000)]{csabai00}
Csabai, I., Connolly, A.\ J., Szalay, A.\ S., \& Budavari, T. 2000,
\aj, 119, 69

\bibitem[da Costa et al.(1991)]{dacosta91}
da Costa, L.\ N., Pellegrini, P.\ S., Davis, M., Meiksin, A., 
Sargent, W.\ L.\ W., \& Tonry, J.\ L.\ 1991, \apjs, 75, 935

\bibitem[da Costa et al.(1998)]{dacosta98}
da Costa, L.\ N., et al.\ 1998, \aj, 116, 1

\bibitem[Davis et al.(1985)]{davis85}
Davis, M., Efstathiou, G., Frenk, C.\ S., \& White,
S.\ D.\ M. 1985, \apj, 292, 371

\bibitem[Davis \& Geller(1976)]{davis76}
Davis, M., \& Geller, M.\ J., 1976, \apj, 208, 13

\bibitem[Davis \& Huchra(1982)]{davis82}
Davis, M., \& Huchra, J.\ P. 1982, \apj, 254, 437

\bibitem[Davis et al.(1988)]{davis88}
Davis, M., Meiksin, A., Strauss, M.\ A., da Costa, L.\ N., \& Yahil, A.
1988, \apj, 333, L9 

\bibitem[Davis, Miller, \& White(1997)]{davis97}
Davis, M., Miller, A., \& White, S.\ D.\ M. 1997, \apj, 490, 63

\bibitem[Davis \& Peebles(1977)]{davis77}
Davis, M., \& Peebles, P.\ J.\ E. 1977, \apjs, 34, 425

\bibitem[Davis \& Peebles(1983)]{davis83}
Davis, M., \& Peebles, P.\ J.\ E. 1983, \apj, 267, 465

\bibitem[de Lapparent, Geller, \& Huchra(1986)]{delapparent86}
de Lapparent, V., Geller, M.\ J., \& Huchra, J.\ P. 1986, \apj, 302, L1

\bibitem[de Lapparent, Geller, \& Huchra(1988)]{delapparent88}
de Lapparent, V., Geller, M.\ J., \& Huchra, J.\ P. 1986, \apj, 332, 44

\bibitem[Diaferio \& Geller(1996)]{diaferio96}
Diaferio, A., \& Geller, M.\ J. 1996, \apj, 467, 19

\bibitem[Dodelson et al.(2001)]{dodelson01}
Dodelson, S., et al. 2001, \apj, submitted,
astro-ph/0107421

\bibitem[Dressler(1980)]{dressler80}
Dressler, A. 1980, \apj, 236, 351

\bibitem[Eisenstein et al.(2001)]{eisenstein01}
Eisenstein, D., et al. 2001, \aj, 122, 2267

\bibitem[Falco et al.(1999)]{falco99}
Falco, E.\ E., et al.
1999, PASP, 111, 438

\bibitem[Feldman, Kaiser \& Peacock(1994)]{feldman94}
Feldman, H., Kaiser, N., \& Peacock, J.\ A. 1994, \apj, 426, 23

\bibitem[Fischer et al.(2000)]{fischer00}
Fischer, P., et al.\ 2000, \aj, 120, 1198

\bibitem[Fisher(1995)]{fisher95a}
Fisher, K.\ B. 1995, \apj, 448, 494

\bibitem[Fisher et al.(1994)]{fisher94}
Fisher, K. B., Davis, M., Strauss, M.\ A., Yahil, A., \& Huchra, J.\ P. 1994,
\mnras, 267, 927

\bibitem[Fisher et al.(1995)]{fisher95} 
Fisher, K.\ B., Huchra, J.\ P., Strauss, M.\ A., Davis, M.,
Yahil, A., \& Schlegel, D. 1995, \apjs, 100, 69

\bibitem[Folkes et al.(1999)]{folkes99}
Folkes, S.\ R., et al. 1999, \mnras, 308, 459

\bibitem[Fry \& Gazta\~naga(1993)]{fry93}
Fry, J.\ N., \& Gazta\~naga, E. 1993, \apj, 413, 447

\bibitem[Fukugita et al.(1996)]{fukugita96}
Fukugita, M., Ichikawa, T., Gunn, J.\ E., Doi, M., Shimasaku, K., \&
Schneider, D.\ P. 1996, \aj, 111, 1748

\bibitem[Geller et al.(1997)]{geller97}
Geller, M.\ J., et al.\ 1997, \aj, 114, 2205

\bibitem[Geller \& Huchra(1989)]{geller89}
Geller, M.\ J., \& Huchra, J.\ P. 1989, Science, 246, 897

\bibitem[Giovanelli \& Haynes(1985)]{giovanelli85}
Giovanelli, R., \& Haynes, M.\ P. 1985, \aj, 90, 2445

\bibitem[Gregory \& Thompson(1978)]{gregory78}
Gregory, S.\ A., \& Thompson, L.\ A. 1978, \apj, 222, 784

\bibitem[Gunn et al.(1998)]{gunn98}
Gunn, J.\ E., Carr, M.\ A., Rockosi, C.\ M., Sekiguchi, M., et al. 1998, 
\aj, 116, 3040

\bibitem[Guzzo et al.(1997)]{guzzo97}
Guzzo, L., Strauss, M.\ A., Fisher, K.\ B., Giovanelli, R., \& Haynes, 
M.\ P. 1997, \apj, 489, 37 

\bibitem[Hamilton(1988)]{hamilton88}
Hamilton, A.\ J.\ S. 1988, \apj, 331, L59

\bibitem[Hamilton(1992)]{hamilton92}
Hamilton, A.\ J.\ S. 1992, \apj, 385, L5

\bibitem[Hamilton(1993)]{hamilton93}
Hamilton, A.\ J.\ S. 1993, \apj, 417, 19

\bibitem[Hamilton(1998)]{hamilton98}
Hamilton, A.\ J.\ S. 1998, ASSL Vol.\ 231: The Evolving Universe, 185

\bibitem[Hatton \& Cole(1998)]{hatton98}
Hatton, S., \& Cole, S. 1998, \mnras, 296, 10

\bibitem[Hermit et al.(1996)]{hermit96}
Hermit, S., Santiago, B.\ X., Lahav, O., Strauss, M.\ A., Davis, M.,
Dressler, A., \& Huchra, J.\ P. 1996, \mnras, 283, 709

\bibitem[Hogg(1999)]{hogg99}
Hogg, D. 1999, astro-ph/9905116

\bibitem[Hubble(1936)]{hubble36}
Hubble, E.\ P. 1936, The Realm of the Nebulae (Oxford University Press: 
Oxford), 79

\bibitem[Huchra et al.(1983)]{huchra83}
Huchra, J.\ P., Davis, M., Latham, D., \& Tonry, J.\ L. 1983, \apjs, 52, 89

\bibitem[Jing, B\"orner, \& Suto(2002)]{jing01}
Jing, Y.\ P., B\"orner, G., \& Suto, Y. 2002, \apj, 564, in press,
astro-ph/0104023

\bibitem[Jing et al.(1998)]{jing98}
Jing, Y.\ P., Mo, H.\ J., \& B\"orner, G. 1998, \apj, 494, 1

\bibitem[Joeveer \& Einasto(1978)]{joeveer78} 
Joeveer, M., \& Einasto, J.\ 1978, IAU Symp.\ 79: Large Scale Structures 
in the Universe, 79, 241

\bibitem[Juszkiewicz, Fisher, \& Szapudi(1998)]{juszkiewicz98}
Juszkiewicz, R., Fisher, K.\ B., \& Szapudi, I.\ 1998, \apjl, 504, L1

\bibitem[Kaiser(1984)]{kaiser84}
Kaiser, N. 1984, \apj, 294, L9

\bibitem[Kaiser(1987)]{kaiser87}
Kaiser, N. 1987, \mnras, 227, 1

\bibitem[Katz, Hernquist, \& Weinberg(1992)]{katz92}
Katz, N., Hernquist, L., \& Weinberg, D.\ H. 1992, \apj, 399, L109

\bibitem[Kauffman et al.(1999)]{kauffmann99}
Kauffmann, G., Colberg, J.\ M., Diaferio, A., \& White, S.\ D.\ M. 1999,
\mnras, 303, 188

\bibitem[Landy, Szalay, \& Broadhurst(1998)]{landy98}
Landy, S.\ D., Szalay, A.\ S., \& Broadhurst, T.\ J. 1998, \apj, 494, L133

\bibitem[Landy \& Szalay(1993)]{landy93}
Landy, S.\ D., \& Szalay, A.\ S. 1993, \apj, 412, 64

\bibitem[Lawrence et al.(1999)]{lawrence99}
Lawrence, A., et al.
1999, \mnras, 308, 897

\bibitem[Loveday et al.(1995)]{loveday95}
Loveday, J., Maddox, S.\ J., Efstathiou, G., \& Peterson, B.\ A. 1995,
\apj, 442, 457

\bibitem[Loveday et al.(1996)]{loveday96}
Loveday, J., Peterson, B.\ A., Maddox, S.\ J., \& Efstathiou, G.\ 1996,
\apjs, 107, 201

\bibitem[Lupton (1993)]{lupton93}
Lupton, R.\ H. 1993, Statistics in Theory and Practice 
(Princeton: Princeton University Press)

\bibitem[Lupton et al.(2001)]{lupton01a}
Lupton, R., Gunn, J.\ E., Ivezi\'c, Z., Knapp, G.\ R., Kent, S., \& Yasuda, N.
2001, in  ASP Conf. Ser. 238, Astronomical Data Analysis Software and
Systems X, ed. F.\ R. Harnden, Jr., F.\ A.\ Primini, and H.\ E.\ Payne (San
Francisco: Astr.\ Soc.\ Pac.), 269

\bibitem[Lupton et al.(2002)]{lupton01}
Lupton, R.\ H., et al. 2002, in preparation

\bibitem[Ma \& Fry(2000)]{ma00}
Ma, C., \& Fry, J.\ N. 2000, \apj, 543, 503

\bibitem[Mann, Peacock, \& Heavens(1998)]{mann98}
Mann, R.\ G., Peacock, J.\ A., \& Heavens, A.\ F. 1998, \mnras, 293, 209

\bibitem[Marzke et al.(1995)]{marzke95}
Marzke, R.\ O., Geller, M.\ J., da Costa, L.\ N., \& Huchra, J.\ P. 1995, 
\aj, 110, 477

\bibitem[McKay et al.(2001)]{mckay01}
McKay, T., et al. 2001, \aj, submitted,
astro-ph/0108013

\bibitem[Melott \& Fry(1986)]{melott86}
Melott, A.\ L., \&  Fry, J.\ N. 1986, \apj, 305, 1

\bibitem[Mo, Jing \& B\"orner(1993)]{mo93}
Mo, H.\ J., Jing, Y.\ P., \& B\"orner, G. 1993, \mnras, 264, 825

\bibitem[Mo, Jing \& B\"orner(1997)]{mo97}
Mo, H.\ J., Jing, Y.\ P., \& B\"orner, G. 1997, \mnras, 286, 979

\bibitem[Mo, McGaugh \& Bothun(1994)]{mo94}
Mo, H.\ J., McGaugh, S.\ S., \& Bothun G.\ D. 1994, \mnras, 267, 129

\bibitem[Narayanan et al.(2000)]{narayanan00}
Narayanan, V.\ K., Berlind, A.\ A., \& Weinberg, D.\ H. 2000, \apj, 528, 1

\bibitem[Norberg et al.(2001)]{norberg01}
Norberg, P., et al. 2001, \mnras, 328, 64

\bibitem[Park et al.(1994)]{park94}
Park, C., Vogeley, M.\ S., Geller, M.\ J., \& Huchra, J.\ P. 1994,
\apj, 431, 569

\bibitem[Peacock et al.(2001)]{peacock01}  
Peacock, J.\ A.\ et al.\ 2001, Nature, 410, 169

\bibitem[Peacock \& Smith(2000)]{peacock00}
Peacock, J.\ A., \& Smith, R.\ E. 2000, \mnras, 318, 1144

\bibitem[Pearce et al.(1999)]{pearce99}
Pearce, F.\ R., et al.
1999, \apj, 521, L99

\bibitem[Peebles(1976)]{peebles76}
Peebles, P.\ J.\ E. 1976, Ap\&SS, 45, 3

\bibitem[Peebles(1980)]{peebles80}
Peebles, P.\ J.\ E. 1980, The Large Scale Structure of the
Universe (Princeton: Princeton University Press)

\bibitem[Percival et al.(2001)]{percival01}
Percival, W.\ J., et al. 2001, \mnras, 327, 1297

\bibitem[Petrosian(1976)]{petrosian76}
Petrosian, V. 1976, \apj, 209, L1

\bibitem[Pier et al.(2002)]{pier01}
Pier, J.\ R., et al. 2002, in preparation

\bibitem[Ratcliffe et al.(1998)]{ratcliffe98}
Ratcliffe, A., Shanks, T., Parker, Q.\ A., Broadbent, A., Watson, F.\ G.,
Oates, A.\ P., Collins, C.\ A., \& Fong, R.\ 1998, \mnras, 300, 417

\bibitem[Regos \& Geller(1991)]{regos91}
Regos, E., \& Geller, M.\ J. 1991, 377, 14

\bibitem[Santiago et al.(1995)]{santiago95}
Santiago, B.\ X., Strauss, M.\ A., Lahav, O., Davis, M., 
Dressler, A., \& Huchra, J.\ P. 1995, \apj, 446, 457

\bibitem[Sargent \& Turner(1977)]{sargent77}
Sargent, W.\ L.\ W., \& Turner, E.\ L. 1977, \apj, 212, L3

\bibitem[Saunders et al.(2001)]{saunders01}
Saunders, W., et al. 2001, \mnras, 317, 55

\bibitem[Schechter(1976)]{schechter76}
Schechter, P. 1976, \apj, 203, 297

\bibitem[Scherrer \& Weinberg(1998)]{scherrer98}
Scherrer, R.\ J., \& Weinberg, D.\ H. 1998, \apj, 504, 607

\bibitem[Schlegel et al.(1998)]{schlegel98}
Schlegel, D.\ J., Finkbeiner, D.\ P., \& Davis, M. 1998, \apj, 500, 525

\bibitem[Schneider et al.(2002)]{richards01}
Schneider, D.\ P., et al.\ 2002, \aj, in press, astro-ph/0110629

\bibitem[Scoccimarro \& Sheth(2002)]{scocc01}
Scoccimarro, R., \& Sheth, R.\ K. 2002, \mnras, 329, 629

\bibitem[Scoccimarro et al.(2001)]{scoccimarro01}
Scoccimarro, R., Sheth, R.\ K., Hui, L., \& Jain, B. 2001, \apj, 546, 20

\bibitem[Scranton et al.(2001)]{scranton01}
Scranton, R., et al. 2001, \apj, submitted,
astro-ph/0107416

\bibitem[Seljak(2000)]{seljak00}
Seljak, U. 2000, \mnras, 318, 203

\bibitem[Shectman et al.(1996)]{shectman96}
Shectman, S.\ A., Landy, S.\ D., Oemler, A., Tucker, D.\ L.,
Lin, H., Kirshner, R.\ P., \& Schechter, P.\ L. 1996, \apj, 470, 172

\bibitem[Shepherd et al.(2002)]{shepherd01}
Shepherd, C.\ W., Carlberg, R.\ G., Yee, H.\ K.\ C., Morris, S.\ L.,
Lin, H., Sawicki, M., Hall, P.\ B., \& Patton, D.\ R. 2002, \apj, in press,
astro-ph/0106250

\bibitem[Sheth(1996)]{sheth96}
Sheth, R. 1996, \mnras, 279, 1310

\bibitem[Sheth et al.(2001a)]{sheth01a}
Sheth, R.\ K., Diaferio, A., Hui, L., \& Scoccimarro, R. 2001, \mnras,
326, 463

\bibitem[Sheth et al.(2001b)]{sheth01b}
Sheth, R.\ K., Hui, L., Diaferio, A., \& Scoccimarro, R. 2001, \mnras,
325, 1288

\bibitem[Siegmund et al.(2002)]{siegmund01}
Siegmund, W., et al. 2002, in preparation

\bibitem[Smith et al.(2002)]{tucker01}
Smith, J.\ A., et al. 2002, \aj, in press

\bibitem[Somerville, Davis, \& Primack(1997)]{somerville97}
Somerville, R.\ S., Davis, M., \& Primack, J.\ R. 1997, \apj, 479, 616

\bibitem[Stoughton et al.(2002)]{stoughton01} 
Stoughton, C., et al. 2002, \aj, in press

\bibitem[Strateva et al.(2001)]{strateva01}
Strateva, I., et al. 2001, \aj, 122, 1861

\bibitem[Strauss et al.(1992)]{strauss92}
Strauss, M.\ A., Huchra, J.\ P., Davis, M., Yahil, A.,
Fisher, K.\ B., \& Tonry, J.\ L. 1992, \apjs, 83, 29

\bibitem[Strauss, Ostriker \& Cen(1998)]{strauss98}
Strauss, M.\ A, Ostriker, J., \& Cen, R. 1998, \apj, 494, 20

\bibitem[Strauss et al.(2002)]{strauss01}
Strauss, M.\ A., et al. 2002, \aj, submitted

\bibitem[Szalay et al.(2001)]{szalay01}
Szalay, A.\ S., et al. 2001, \apj, submitted,
astro-ph/0107419

\bibitem[Szapudi et al.(2002)]{szapudi01}
Szapudi, I., et al. 2002, \apj, in press,
astro-ph/0111058

\bibitem[Tegmark et al.(1998)]{tegmark98}
Tegmark, M., Hamilton, A., Strauss, M.\ A., Vogeley, M., \& Szalay, 
A.\ S. 1998, \apj, 499, 555

\bibitem[Tegmark, Taylor, \& Heavens(1997)]{tegmark97}
Tegmark, M., Taylor, A.\ N., \& Heavens, A.\ F. 1997, \apj, 480, 22 

\bibitem[Tegmark et al.(2002)]{tegmark01}
Tegmark, M., et al. 2002, \apj, in press,
astro-ph/0107418

\bibitem[Tucker et al.(1997)]{tucker97}
Tucker, D.\ L., et al.
1997, \mnras, 285, L5  

\bibitem[Uomoto et al.(2002)]{uomoto01}
Uomoto, A., et al.\ 2002, in preparation

\bibitem[Vanden Berk et al.(2002)]{vandenberk01} 
Vanden Berk, D.\ E., et al. 2002, in preparation

\bibitem[van de Weygaert \& van Kampen(1993)]{weygaert93}
van de Weygaert R., \& van Kampen E. 1993, \mnras, 263, 481

\bibitem[Vettolani et al.(1998)]{vettolani98}
Vettolani, G., et al.\ 1998, \aaps, 130, 323

\bibitem[Vogeley \& Szalay(1996)]{vogeley96}
Vogeley, M.\ S., \& Szalay, A.\ S. 1996, \apj, 465, 34

\bibitem[White, Hernquist, \& Springel(2001)]{white01}
White, M., Hernquist, L., \& Springel, V. 2001, \apjl, 550, L129

\bibitem[White, Tully, \& Davis(1988)]{white88}
White, S.\ D.\ M., Tully, R.\ B., \& Davis, M. 1988, \apjl, 333, L45

\bibitem[Willmer, da Costa, \& Pellegrini(1998)]{willmer98}
Willmer, C.\ N.\ A., da Costa, L.\ N., \& Pellegrini, P.\ S. 1998, \aj, 
115, 869

\bibitem[Yasuda et al.(2001)]{yasuda01}
Yasuda, N., et al.\ 2001, \aj, 122, 1104

\bibitem[York et al.(2000)]{york00}
York, D.\ G. et al.\ 2000, \aj, 120, 1579

\bibitem[Yoshikawa et al.(2001)]{yoshikawa01}
Yoshikawa, K., Taruya, A., Jing, Y.\ P., \& Suto, Y. 2001, \apj, 558, 520

\bibitem[Zurek et al.(1994)]{zurek94}
Zurek, W.\ H., Quinn, P.\ J., Salmon, J.\ K., \& Warren, M.\ S. 1994, \apj, 
431, 559

\bibitem[Zwicky(1937)]{zwicky37}
Zwicky, F. 1937, \apj, 86, 217

\bibitem[Zwicky et al.(1968)]{zwicky68}
Zwicky, F., Herzog, E., Wild, P., Karpowicz, M., \& Kowal, C.,
1961-1968, {\it Catalog of Galaxies and of Clusters of Galaxies},
Vols. 1-6, (Pasadena: California Institute of Technology)

\end{thebibliography}
\end{document}